\documentclass[usegraphicx,useAMS,usenatbib]{mn2e}
\usepackage{times}
\usepackage{amssymb}
\usepackage{amsmath}
\usepackage{graphicx}
\usepackage{euscript}
\usepackage{rotating}
\usepackage{epsfig}

\def\hz{{\rm\thinspace Hz}}
\def\mhz{{\rm\thinspace MHz}}
\def\ghz{{\rm\thinspace GHz}}
\def\cmsqps{{\rm\thinspace cm^{2}~s^{-1}}}
\def\kmpspmpc{\hbox{$\rm\thinspace km~s^{-1}~Mpc^{-1}$}}
\def\kev{{\rm\thinspace keV}}
\def\G{{\rm\thinspace G}} 

\begin{document}

\newcommand{\Mpc}{\rm\thinspace Mpc}
\newcommand{\kpc}{\rm\thinspace kpc}
\newcommand{\pc}{\rm\thinspace pc}
\newcommand{\km}{\rm\thinspace km}
\newcommand{\m}{\rm\thinspace m}
\newcommand{\cm}{\rm\thinspace cm}
\newcommand{\cmps}{\hbox{$\cm\s^{-1}\,$}}
\newcommand{\cmpssq}{\hbox{$\cm\s^{-2}\,$}}
\newcommand{\cmsq}{\hbox{$\cm^2\,$}}
\newcommand{\cmcu}{\hbox{$\cm^3\,$}}
\newcommand{\pcmcu}{\hbox{$\cm^{-3}\,$}}
\newcommand{\pcmcuK}{\hbox{$\cm^{-3}\K\,$}}

\newcommand{\yr}{\rm\thinspace yr}
\newcommand{\Gyr}{\rm\thinspace Gyr}
\newcommand{\s}{\rm\thinspace s}
\newcommand{\ks}{\rm\thinspace ks}

\newcommand{\GHz}{\rm\thinspace GHz}
\newcommand{\MHz}{\rm\thinspace MHz}
\newcommand{\Hz}{\rm\thinspace Hz}

\newcommand{\K}{\rm\thinspace K}

\newcommand{\Kpcmc}{\hbox{$\K\cm^{-3}\,$}}

\newcommand{\g}{\rm\thinspace g}
\newcommand{\gpcm}{\hbox{$\g\cm^{-3}\,$}}
\newcommand{\gpcmps}{\hbox{$\g\cm^{-3}\s^{-1}\,$}}
\newcommand{\gps}{\hbox{$\g\s^{-1}\,$}}
\newcommand{\Msun}{\hbox{$\rm\thinspace M_{\odot}$}}
\newcommand{\Msunpc}{\hbox{$\Msun\pc^{-3}\,$}}
\newcommand{\Msunpkpc}{\hbox{$\Msun\kpc^{-1}\,$}}
\newcommand{\Msunppc}{\hbox{$\Msun\pc^{-3}\,$}}
\newcommand{\Msunppcpyr}{\hbox{$\Msun\pc^{-3}\yr^{-1}\,$}}
\newcommand{\Msunpyr}{\hbox{$\Msun\yr^{-1}\,$}}

\newcommand{\MeV}{\rm\thinspace MeV}
\newcommand{\keV}{\rm\thinspace keV}
\newcommand{\eV}{\rm\thinspace eV}
\newcommand{\erg}{\rm\thinspace erg}
\newcommand{\Jy}{\rm Jy}
\newcommand{\ergpcmc}{\hbox{$\erg\cm^{-3}\,$}}
\newcommand{\ergcmcups}{\hbox{$\erg\cm^3\ps\,$}}
\newcommand{\ergpcmps}{\hbox{$\erg\cm^{-3}\s^{-1}\,$}}
\newcommand{\ergpcmsqps}{\hbox{$\erg\cm^{-2}\s^{-1}\,$}}
\newcommand{\ergpcmsqpspA}{\hbox{$\erg\cm^{-2}\s^{-1}$\AA$^{-1}\,$}}
\newcommand{\ergpcmsqpspsr}{\hbox{$\erg\cm^{-2}\s^{-1}\sr^{-1}\,$}}
\newcommand{\ergpcmcups}{\hbox{$\erg\cm^{-3}\s^{-1}\,$}}
\newcommand{\ergps}{\hbox{$\erg\s^{-1}\,$}}
\newcommand{\ergpspmp}{\hbox{$\erg\s^{-1}\Mpc^{-3}\,$}}
\newcommand{\keVpcmsqpspsr}{\hbox{$\keV\cm^{-2}\s^{-1}\sr^{-1}\,$}}

\newcommand{\dyn}{\rm\thinspace dyn}
\newcommand{\dynpcmsq}{\hbox{$\dyn\cm^{-2}\,$}}

\newcommand{\kmps}{\hbox{$\km\s^{-1}\,$}}
\newcommand{\kmpspmp}{\hbox{$\km\s^{-1}\Mpc{-1}\,$}}
\newcommand{\kmpspMpc}{\hbox{$\kmps\Mpc^{-1}$}}

\newcommand{\Lsun}{\hbox{$\rm\thinspace L_{\odot}$}}
\newcommand{\Lsunppc}{\hbox{$\Lsun\pc^{-3}\,$}}

\newcommand{\Zsun}{\hbox{$\rm\thinspace Z_{\odot}$}}
\newcommand{\gauss}{\rm\thinspace gauss}
\newcommand{\arcsecond}{\rm\thinspace arcsec}
\newcommand{\chisq}{\hbox{$\chi^2$}}
\newcommand{\delchi}{\hbox{$\Delta\chi$}}
\newcommand{\ph}{\rm\thinspace ph}
\newcommand{\sr}{\rm\thinspace sr}

\newcommand{\pccm}{\hbox{$\cm^{-3}\,$}}
\newcommand{\psqcm}{\hbox{$\cm^{-2}\,$}}
\newcommand{\pcmsq}{\hbox{$\cm^{-2}\,$}}
\newcommand{\pmpc}{\hbox{$\Mpc^{-1}\,$}}
\newcommand{\pmpccu}{\hbox{$\Mpc^{-3}\,$}}
\newcommand{\ps}{\hbox{$\s^{-1}\,$}}
\newcommand{\pHz}{\hbox{$\Hz^{-1}\,$}}
\newcommand{\pcmK}{\hbox{$\cm^{-3}\K$}}
\newcommand{\phpcmsqps}{\hbox{$\ph\cm^{-2}\s^{-1}\,$}}
\newcommand{\psr}{\hbox{$\sr^{-1}\,$}}
\newcommand{\pspsqas}{\hbox{$\s^{-1}\,\arcsecond^{-2}\,$}}

\newcommand{\ergpspcmpK}{\hbox{$\erg\s^{-1}\cm^{-1}\K^{-1}\,$}}

\title{Particle Energies and Filling Fractions of Radio Bubbles in
  Cluster Cores} \author[Dunn \& Fabian]
{\parbox[]{6.in} {R.J.H. Dunn \thanks{E-mail: rjhd2@ast.cam.ac.uk} and A.C. Fabian \\
    \footnotesize
    Institute of Astronomy, Madingley Road, Cambridge CB3 0HA\\
  }}

\maketitle

\begin{abstract}
Using \emph{Chandra} images of cluster cores with clear radio bubbles, we
have determined $k$, which is the ratio of the total particle energy to
that of the electrons radiating between $10\mhz$ and $10\ghz$. Radiative
and dynamical constraints on the bubbles indicate that the ratio of
the energy factor, $k$, to the volume filling factor, $f$, lies within
the range $1 \lesssim k/f \lesssim 1000$.  Assuming pressure
equilibrium between the radio-emitting plasma and the surrounding
X-ray gas, none of the lobes have equipartition between 
relativistic particles and magnetic field. There is no evidence for
any dependence of the upper limit of the $k/f$ ratio on any physical parameter of the
cluster or the radio source.  The distribution of the upper limit on $k/f$ appears to be
bimodal, the value for some clusters being $\sim 3$ and for the others
$\sim 300$.  We show that this is may due to the composition of the
jet which forms the bubbles, the variation in the volume filling
fraction or variation in the amount of re-acceleration occurring in the bubble.

\end{abstract}


\section{Introduction}

Radio lobes which emit synchrotron radiation contain relativistic electrons and
magnetic fields.  Disentangling the pressures of each component is
difficult, and traditionally it has been assumed that there is
equipartition between the particles and the field \citep{Burbidge},
which corresponds closely to the minimum energy condition.
Now, however, when the bubbles are embedded in an Intra Cluster Medium (ICM),
this degeneracy can be removed by
measuring the thermal pressure of the ICM and assuming
pressure equilibrium between the bubbles and the surrounding X-ray
gas as well as the lack of strong shocks.  In particular, FR I
sources in low redshift clusters imaged with \emph{Chandra} often show
holes in the X-ray emission coinciding with the radio lobes
(e.g. Hydra A, \citep{McNamaraHydra00}; Perseus, \citep{ACF_complex_PER00}; A2052,
\citep{Blanton01}; A2199, \citep{JohnstoneA2199}; Centaurus, \citep{SandersCent02}),
the first of which was discovered in the Perseus cluster with
\emph{ROSAT} \citep{Bohringer}.  A recent compilation is given by \citet{Birzan04}.

Here we perform a detailed study of a sample of clusters which have
clear radio bubbles.  Following the approach first detailed in
\citet{Celotti02}, we determine $k/f$, where $k$ is the
ratio of the total relativistic particle energy to that in electrons
emitting synchrotron radiation between $10\mhz$ and $10\ghz$, and $f$ is the
volume filling factor of the relativistic plasma in the bubble.

\citet{Celotti02} calculated $k/f$ for the northern radio bubble in the
Perseus cluster and obtained a value of $180 < k/f < 500$.  We find a
value that is in broad agreement with theirs, and find similar values for
M84 and PKS 1404-267.  However the remaining clusters in our sample,
including A2052, A2199, A4059, Centaurus and Hydra A, appear to have
much lower values.  $k/f$ appears to be either in the range $\sim1-10$ or $\sim100-1000$.  We
discuss the implications of this result on the evolution and formation
of the bubbles.  We use $H_0=70\kmpspmpc$ throughout.

\section{Data Analysis}\label{Dataanal}

We use standard synchrotron theory to quantify the properties of the
particles present in the bubbles, following the analysis presented in
\citet{Celotti02}.  We give below some of the formulae required in
order to define the variables.  The total energy in relativistic electrons radiating between
 $\nu_{1}$ to $\nu_{2}$, with a spectral index $\alpha$
($S(\nu) \propto \nu^{\alpha}$), in a magnetic field
$B$, producing a flux density $S_{\nu}$ at $\nu$, is
\begin{eqnarray}
 E_{\rm e}\!\!\!\!&=&\!\!\!\!4\pi\times10^{12}\Big(\frac{cz}{H_0}\Big)^2\Big(1+\frac{z}{2}\Big)^2
 \frac{S_{\nu}}{\nu^{\alpha}}
 \frac{\nu^{0.5+\alpha}_{2}-\nu^{0.5+\alpha}_{1}}{\alpha+0.5}
 B^{-3/2} \nonumber\\
 \!&\approx&\!aB^{-3/2} \erg , \label{Eparteqn}
\end{eqnarray}
\noindent
where  $H_0= 70\kmpspmpc$.  
It has been assumed that the particle energy distribution extends from
$\nu_1 = 10\mhz$ to $\nu_2 = 10\ghz$ as the radio images analysed
were taken at frequencies between $1$ and $8\ghz$.  Therefore the total
energy in particles and magnetic field is 
\[
 E_{\rm tot} = kE_{\rm e} + Vf\displaystyle\frac{B^{2}}{8\pi} =
 akB^{-3/2} + bf{B^{2}} \erg ,
\]
\noindent
where $V$ is the volume of the bubble, and $f$ represents the volume
filling factor of the relativistic plasma.  $k$ accounts for the
additional energy from relativistic particles accompanying the
electrons that radiate above $10\mhz$ and any non-relativistic
component ($k=1$ for an electron-positron plasma emitting only in the
above waveband; a typical value used in the literature is $k=100$).
The bubbles are allowed to be aspherical, where $r_{\rm l}$
(``length'') is the radius along the
jet direction and $r_{\rm w}$ (``width'') is the radius across the jet
direction, where a symmetry axis has been assumed along the jet, hence
the volume is $V=4\pi r_{\rm l} r_{\rm w}^2/3$.

With  the condition that there is simple equipartition between the
energy present in particles and that present in the magnetic field, the magnetic field strength is
\[
 B_{\rm eq} = \Big(\displaystyle\frac{a}{b}\Big)^{2/7}\Big(\displaystyle\frac{k}{f}\Big)^{2/7}\G.
\]
\noindent
For the minimum energy magnetic field, replace $(a/b)^{2/7}$ with $(3a/4b)^{2/7}$ in the
above equation.

If the relativistic gas is in equilibrium at equipartition with the
thermal pressure from the gas in the rims surrounding the
radio lobes ($P_{\rm th}$), then the equipartition value, $k/f_{\rm eq}$, can be obtained.

From now on we assume that the plasma is not in equipartition but
 still in pressure
equilibrium, hence $k/f$ can be found from
\begin{equation} \label{k/f}
\displaystyle \frac{k}{f} = \Big(P_{\rm th} - \frac{B^2}{8\pi}\Big)\frac{3V}{a}B^{3/2}.
\end{equation}
\noindent
To calculate $k/f$ from Equation \ref{k/f}, estimates
of the magnetic field strength in the bubbles are obtained from
radiative and dynamical constraints.

A minimum value 
of $k/f=1$ can be understood for an electron-positron plasma which
fills all of the bubble.  However, an extremal value can be determined by differentiating
Equation \ref{k/f}, and finding the corresponding magnetic field gives
a maximum value, $k/f_{\rm max}$, which is $50 \%$ greater than the equipartition
value.  The absolute maximum value for the magnetic field is when $k/f = 1$ from when
\mbox{$(P_{\rm th}-B^2/8\pi)\sim 0$,}
which is 1.53 times greater than the magnetic field at $k/f_{\rm max}$, and 1.15
times the equipartition magnetic field.  The field
for $k/f_{\rm max}$ is the limit up to which the $B^2$ term can be
ignored.  Any further increase in $B$ and this term becomes dominant
and $k/f$ decreases until it equals zero.  After that the magnetic
pressure is such that the bubble would be over pressured, even with
$k=1$, and for pressure equilibrium fewer particles than observed
would be required.

A limit on the magnetic field present in the bubbles can be deduced
from the fact that GHz radio emission is seen throughout the
bubbles. Therefore the synchrotron cooling time of the relativistic
electrons present is
\begin{equation}\label{synctime}
 t_{\rm sync} = 2.7 \times 10^{7} B^{-3/2}_{-5} \nu^{-1/2}_{9}\yr,
\end{equation}
\noindent
where $B=10^{-5}B_{-5}\G$
and $\nu=10^{9}\nu_{9}\hz$, has to be greater than the age of the
bubble, assuming that there is no re-acceleration of the electrons in
the bubble (for further discussion see Section \ref{Reacc}).  However Equation \ref{synctime} is
only valid for strong enough magnetic fields (for further discussion
see Section \ref{timescales}).

The age of the bubble can be determined
from the fact that the X-ray rims surrounding the radio bubbles appear not to be strongly
shocked, hence they must be expanding at less than the sound speed of the
gas in the rims, $ c_{\rm s} = \sqrt{\gamma k T/\mu m_{\rm{H}}}$, where
$\gamma$ is the ratio of the heat capacities and $\mu = 0.62$.  And therefore the age of the
bubbles must be greater than $ t_{\rm sound} = 2r_{\rm l}/c_s $ where
$r_{\rm l}$ is the radius of the bubble.  Twice the radius has been used
as the bubble is blown from the radio source at the centre of the
galaxy, which is at one edge of the bubble, rather than the centre,
and so the ``front edge'' travels twice the bubbles' radius.  This edge has
travelled furthest during the lifetime of the bubble and so provides a
limit on its age.

If the bubbles rise upwards at their buoyancy velocity, $ v_{\rm b} =
\sqrt{2gV/SC_{\rm D}} $, 
where $S$ is the cross-sectional area of the bubble, $V$ is the
volume, $g = GM(<R_{\rm dist})/R_{\rm dist}^2$ for the bubble (centre) being at $R_{\rm dist}$ from the
cluster core and $C_{\rm D} = 0.75$ is the drag coefficient
\citep{Churazov01}, then the
age of the bubble can be estimated as $t_{\rm buoy} = R_{\rm
  dist}/v_{\rm b}$, the travel time to their current position.  The enclosed mass was estimated from cluster mass profiles or using a
linear interpolation from the Abell radius and mass as listed in
\citet{Reiprich}.  The latter is probably only correct to a factor of a few, though
the uncertainty was taken into account in the calculation.  The age of
the bubble can also be estimated from the time required to
refill the displaced volume as the bubble rises upward
\citep{McNamaraHydra00}, $t_{\rm refill} = 2R_{\rm dist}\sqrt {r/GM(<R_{\rm dist})}$.

As the synchrotron lifetime of the GHz electrons must be greater than
the age of the bubble, limits can be placed on the magnetic
field, assuming that $\nu_9 = 1$.  

The value obtained shows whether the equipartition solution is possible and
allows $k/f$ to be calculated from Equation \ref{k/f}.  The limits
obtained for $k/f$ may be higher than the one predicted for
equipartition.  However, the equipartition condition is determined from the magnetic field strength,
and Fig 7. of \citet{Celotti02} shows that
it is possible that limits on $k/f$ are larger than the equipartition
value even though the magnetic
field is less than its equipartition value.

In the Perseus cluster there are weak shocks
visible in the X-ray images \citep{ACF_complex_PER00}.  Assuming that these have been created by
the earlier supersonic expansion of the bubble, then a maximum
age for the bubble can be estimated.  Currently
these shocks are travelling at the local sound speed,
however, in the past the bubble that caused them was travelling
supersonically, and so the average speed of the shocks over the bubble
lifetime will be greater than the sound speed.  Hence the age of the
bubble must be \emph{less} than $ t_{\rm shock} = R_{\rm shock}/c_{\rm s}$
where $R_{\rm shock}$ is the shock radius, centred on the central
radio source.  

Although this is an upper limit on the age of the
bubble, there is no requirement that the synchrotron lifetime of the
electrons is \emph{less} than this value.  However, in the assumption
that it is, then this upper age limit, gives a lower limit on the
magnetic field, and hence a lower limit on
the value for $k/f$.  In the Perseus cluster, which is the only
cluster in the sample with a shock front visible, this limit together with the value
obtained from the condition that there are no strong shocks at the
rims gives the lowest range of $k/f$ that can be understood from
physical arguments.

\citet{ACF_Halpha_PER03} showed that some of the H$\alpha$ filaments seen in
the Perseus cluster could be interpreted as streamlines behind the
western ``ghost'' bubble.  This indicated that the flow is laminar, and implies a
Reynolds number of less than 1000.  Using this, a value of $4
\times 10^{27} {\rm\thinspace cm^{2}~s^{-1}}$ was obtained for the kinematic
viscosity.  The Reynolds number, $Re$, was calculated for each bubble analysed,
assuming the viscosity was the above value, and also a lower bound on
the viscosity was calculated from the limit on the Reynolds number of
1000 assuming that the flow is laminar in each cluster.

\subsection{Non-uniform Magnetic Fields}\label{nonuniform}

It is unlikely that the magnetic fields present in the bubbles are
completely uniform over the entire bubble.  The effect of non-uniform
magnetic fields was investigated very simply, using a centrally peaked
linear distribution ($B \propto -|r|$).  This gave limits on $k/f$ that
were a factor of two greater than if a uniform magnetic field was used,
for the case of Perseus (Table \ref{Nonunif}).  It was checked that the calculation used
for the non-uniform field produced the same limits on $k/f$ for the a
uniform field as were produced from the other calculations.

\begin{table} \centering \caption{\scshape \label{Nonunif}: Effect of
    Non-uniform Magnetic Field on $k/f$}
\begin{tabular} {l l l l l}
\hline
\hline
Cluster Lobe & Timescale & Uniform Field & Non-uniform Field \\

\hline
Perseus North & Sound & 573  & 1083 \\
              & Shock & 390  &  735 \\ 
Perseus South & Sound & 1025 &  1937 \\
              & Shock & 762  &  1436 \\
\hline
\end{tabular}
\begin{quote}
{\scshape Notes:}\\
The limits on $k/f$ for the sound speed and shock calculations from
the Perseus Cluster with a
uniform and non-uniform magnetic field.
\end{quote}
\end{table}

\subsection{Energetics}

The energies and powers were also calculated, using $E=PV$ for
the energy, and $\mathcal{P}=E/t$ for the powers using all of the timescales
calculated above.  However, if the energy required to expand the
bubble as well as the energy in the bubble is accounted for, then, for
slow expansion rates:
\[
E = \frac{1}{\gamma-1} PV + P{\rm dV} \approx \frac{\gamma}{\gamma-1} PV,
\]
\noindent
where $V$ is the volume of the bubble and $\gamma$ is the mean
adiabatic index of the fluid in the bubble (5/3 for non-relativistic
gas or 4/3 for relativistic gas).  For the relativistic case (assumed
as there is synchrotron emission in the cavity) this increases the
energy by a factor of 4 (5/2 for non-relativistic), and hence also the
power.  Therefore the values for the energy in Table \ref{kftable} should be
multiplied by the appropriate $\gamma/(\gamma-1)$.

\section{Dynamical Constraints}\label{DynCons}

\begin{figure}
\includegraphics[width=1.0 \columnwidth]{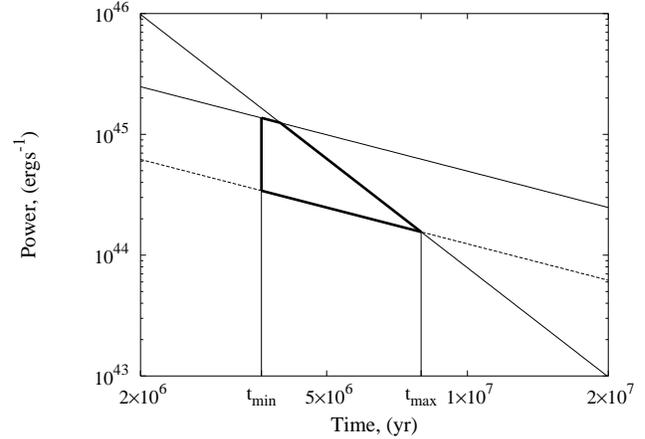}

\caption  {\label{Dynconst} \small{The dynamical constraints given in
    Section \ref{DynCons} shown for the northern Perseus bubble.  The
    vertical lines show the minimum $t_{\rm min}$, obtained from the
    combination of the energy and sound speed equations (\ref{A}) and (\ref{B}), and maximum $t_{\rm
    max}$ ages of the bubble, obtained from the combination of the
    energy and buoyancy equations (\ref{A}) and (\ref{C}).  The inclined lines show the limits on
    the power of the source,
    the upper shallow line is from Equation \ref{A} and the steep line
    is from Equation \ref{lpropt3}.  The dotted shallow line
    represents the minimum power required to move the gas to make
    the hole ($P_{\rm th}V/t$).  In all the equations, $f$ has been taken
    as 1, and $P_{\rm th}$ is the value obtained from the temperature and
    density of the bubbles.}}
\end{figure}

$P{\rm dV}$ work is done by the bubble on its surroundings during its
expansion, equating to a power $\mathcal{P}\approx P_{\rm th}fV/t$.  The jet must be
powerful enough to inflate the holes, and an estimation of this power
can be obtained by considering the evolution of the expanding bubble
in a medium of constant pressure \citep{Churazov01, Churazov00}.  
We modify their expressions for the expansion of the bubble to include
the effect of gas clumping through the volume filling factor.  From
the internal energy of the hole and the compression of the surrounding
gas during subsonic expansion

\begin{equation}\label{A}
 \displaystyle \frac{\mathcal{P}t}{P_{\rm th}f} = V
 \frac{\gamma}{\gamma -1} = {\rm F},
\end{equation}
\noindent
where $\gamma$ for the radio-emitting plasma is assumed
to be 4/3.  Secondly, from the condition that there are no strong
shocks at the bubble rims, the expansion rate must be subsonic, and so
\begin{equation}\label{B}
 \displaystyle \frac{\mathcal{P}}{P_{\rm th}ft^{2}} < 36 \pi c_{\rm
 s}^3 \frac{\gamma}{\gamma -1} = {\rm G}.
\end{equation}
\noindent
Finally, assuming that the bubble has not detached from the source,
which seems to be the case for the majority of the clusters, then its
velocity is less than the buoyancy velocity, giving
\begin{eqnarray}\label{C}
\mathcal{P}&=&\displaystyle \frac{36 \pi \gamma}{\gamma -1} \Big(
\frac{8}{3C_{\rm D}}
\Big)^{3/2} \Big( \frac{r}{R} \Big)^{3/2} \Big( \frac{GM(<R)}{R}
\Big)^{3/2} P_{\rm th} t^2 f^{5/2} \nonumber\\ 
&=&{\rm H} P_{\rm th} t^2 f^{5/2}.
\end{eqnarray}
Combining these expressions produces other limits on the ages
of the radio bubbles --- (\ref{A}) and (\ref{B}) produce a minimum
age for the hole, $t_{\rm min}=(\frac{{\rm F}}{{\rm G}})^{1/3}$; and (\ref{A}) and
(\ref{C}) lead to a relation
\begin{equation} \label{lpropt3}
\mathcal{P}<{\rm H}^{-2/3}{\rm F}^{5/3}P_{\rm th}t^{-3}.
\end{equation}
\noindent
There is also a minimum power for the source, from $P_{th}V/t$
for $f=1$.  The intersection of this with (\ref{lpropt3}) gives the
maximum age of the hole as
\[
 t_{\rm max} = \sqrt{ \frac{{\rm F}^{5/3}{\rm H}^{-2/3}}{V}}.
\]
\noindent
These relations are only valid for subsonic expansion.  For Perseus at
least, there exists a weak shock indicating that there has been
supersonic expansion in the past \citep{ACF_deep_PER03}.  Therefore the agreement between the
timescales given by these relations, with the ones drawn from the
other physical arguments may give a crude indication of what proportion of
its lifetime the bubble has spent in subsonic expansion.  The allowed region
for the bubble from these constraints in the Power--Age plot in Fig. \ref{Dynconst} is
the highlighted polygon defined by the minimum and maximum ages, the power
limits from Equations \ref{A} and \ref{lpropt3} and the minimum
power for the source. 

\section{Results}\label{results}

The individual source properties are listed in Table \ref{source
  prop}.  In A2199, the inner jet-like feature visible on the radio overlays in
\citet{JohnstonePKS} has small X-ray depressions associated with the
knots at the ends of the jet (Inner East (I-E) and Inner West (I-W)), and these were analysed separately from the
extended primary emission.  If the source has prominent X-ray rims
around the radio bubbles, e.g. Perseus, then these have been used to
define the size of the bubbles.  Otherwise the spatial extent of the
radio emission has been used to define the size of the radio
bubble.  In some cases the GHz radio
emission does not entirely fill the X-ray defined holes - e.g. A2052.  In Cygnus A, the fluxes used in the calculation do not
include the hotspots at the end of the jet, to reduce the effect of
re-acceleration of electrons (for further discussion see Section \ref{Reacc}).

\begin{table*} 
\caption{\scshape \label{source prop}: Source Properties}
\begin{tabular}{l l l c r r r r c r r r c}
\hline
\hline
Cluster & Lobe$^{(1)}$ & Redshift & $\alpha$ & log$(P_{5\ghz})$ & $R_{\rm
  dist}^{(2)}$
& $r_{\rm l}$ & $r_{\rm w}$ & $R_{\rm shock}$&
$M_{\rm encl}$ & $kT$ & $n_{\rm e}$ & References\\
& & & & (W/Hz)& (kpc) & (kpc) & (kpc) &  (kpc) & ($10^{12} M_{\odot})$ &
(eV) & (${\rm cm}^{-3}$)&\\
\hline
A133 & Rel, R &0.054  &$-2.0\pm 1.0$&  23.0&  40.0& 50.0 &20.0  & - & 8.0& 2400 & 0.01& 1,2,3,4,5,6,7   \\
\\
A2052 & N, X & 0.035 & $-1.6\pm 0.3$& 24.7 & 10.0 &10.0  & 10.0 & - & 4.0& 1100 & 0.049&  7,8,9,10,11,12,13,14   \\
 & S, X&  & &  & 15.0 & 15.0 & 15.0 & - &5.0 & 1100 & 0.049&   \\
\\
A2199 & E, R  & 0.031 &$-1.9\pm0.3$ &  24.3& 20.2 & 16.9 & 12.2 & - & 1.6& 3100 & 0.017& 13,15,16,17  \\
 & W, R &  & &  & 21.5  & 20.9 & 14.4 &-  &1.7 & 3200 & 0.017&     \\
 & I-E, R &  & &  & 4.23 &  1.0& 1.0 & - & 0.08&  2050& 0.056&    \\
 & I-W, R &  & &  & 4.23 & 1.5 & 1.5&-  & 0.08  &  2050& 0.056&    \\
\\
A4059 & N, R &0.050  &$-1.4\pm0.3$ & 24.0 & 5.44 &5.44  & 2.98 & - & 1.5&  2000& 0.028&7,13,14,18,19  \\
 &S, R  &  & &  & 3.92 & 3.92 & 4.48 & - &1.1 & 2000 & 0.028&     \\
\\
Centaurus & E, R &  0.010&$-1.3\pm0.2$ &25.8  & 1.45 &1.45 & 1.13 &-  &0.020 &  800& 0.09&13,20     \\
 &E, R  &  & &  & 2.00 & 1.88 & 1.02 & - & 0.025&  1000& 0.08&     \\
\\
Cygnus A & E, R & 0.056 & $-1.0\pm0.3$& 27.7 & 44.1 &30.3  & 23.7 & - & 3.0&  4000& 0.05& 13,21,22,23,24    \\
 & W, R &  & &  & 48.0 & 33.9 &  25.2&-  & 5.0&  5500& 0.04&     \\
\\
Hydra A & N, R &0.052  &$-1.2\pm0.4$ & 26.3 & 17.8 & 20.0 & 8.5 & - &2.0&  2900& 0.041&13,25,26,27    \\
 & S, R &  & &  &  22.2& 16.7 & 6.9 &-  & 2.0&  3000& 0.036&     \\
\\
M84 &N, R  &0.0035  &$-0.51\pm0.02$ & 22.0 & 2.7 & 2.1 & 2.5 & - & 0.2&  650& 0.04& 28,29,30    \\
 &S, R  &  & &  &  3.8&  4.0&  2.6& - &  0.2 & 650&  0.04&    \\ 
\\
M87 & Bud, R & 0.0044 &$-1.2\pm0.4$ & 24.4 &  3.11&  1.31&  1.31& - & 0.14&  1500& 0.20&13,31    \\
\\
MKW3s & Rel, R &  0.045& $-2.7\pm0.05$& 23.8 & 56.4 & 12.9 & 17.9 &- & 16.0 & 3200& 0.004& 7,32,33  \\
\\
Perseus & N, X &0.018  & $-1.0\pm0.2$& 25.9 & 8.15 & 8.15 & 8.15 & 24.0 &0.24 &  2604& 0.064& 13,34,35,36,37,38    \\
 & S, X &  & &  & 8.89 &  8.89&  8.89& 24.0 & 0.26&  3175& 0.055&     \\
\\
PKS 1404 & E, X &0.022  & $-0.39\pm0.03$& 23.8 & 3.94 &  2.29&1.58  & - &0.40&  930& 0.056& 39,40,41,42  \\
 & W, X &  & &  & 2.75 &  1.33& 1.7 & - & 0.40&  820& 0.066&     \\

\hline
\end{tabular}
\begin{quote} {\scshape References:}\\
(1) \citet{FujitaSarazin02};
(2) \citet{Slee01};
(3) \citet{Owen&Ledlow97};
(4) \citet{Rizza99};
(5) \citet{SleeReynolds84};
(6) \citet{KomissarovGubanov94};
(7) \citet{Reiprich};
(8) \citet{Blanton01}; 
(9) \citet{Blanton03};
(10) \citet{Burns90};
(11) \citet{Stefanachi02};
(12) \citet{Zhao93};
(13) \citet{TaylorCent02};
(14) G. Taylor, private comm. (2004);
(15) \citet{JohnstoneA2199}; 
(16) \citet{Andernach}; 
(17) \citet{Burns83}
(18) \citet{ChoiA4059};
(19) \citet{TaylorA4059};
(20) \citet{SandersCent02};
(21) \citet{PerleyCygnusA}; 
(22) \citet{Carilli91};
(23) \citet{SmithCygnusA2002};
(24) \citet{Alexander84};
(25) \citet{McNamaraHydra00}; 
(26) \citet{TaylorHydra90}; 
(27) \citet{Voigt04}; 
(28) \citet{FinoguenovJones02};
(29) \citet{Terasranta01};
(30) \citet{CaoRawlings04};
(31) \citet{FormanM87}; 
(32) \citet{Mazzotta}; 
(33) \citet{McNamara90}; 
(34) \citet{Celotti02};
(35) \citet{ACF_complex_PER00}; 
(36) \citet{ACF_Halpha_PER03}; 
(37) \citet{Voigt04};
(38) \citet{Pedlar90}; 
(39) \citet{Drinkwater}; 
(40) \citet{JohnstonePKS};
(41) \citet{BrownBurns91};
(42) \citet{Bettoni03}\\

{\scshape Notes:}\\
(1) The codes for the Lobes are N---Northern, S---Southern, E---Eastern,
    W---Western, X---sizes from X-ray image, R---sizes from Radio image,
    I-E/W---Inner lobes in A2199, Bud---as described in
    \citet{FormanM87}, Rel---Relic source (for A133 as described in
    \citet{FujitaSarazin02}).\\
(2) All the values given in the above table except the radio power
have an uncertainty associated with them.  Except for
the spectral index, they are not quoted as they have limited effect on
the calculated values.  The effect of the uncertainties in $\alpha$ on
the resultant uncertainties in $k/f$ is
large and so are stated here.  For further discussion see text.

\end{quote}
\end{table*}

The limits on $k/f$ calculated from the different timescales are tabulated in Table \ref{kftable}.  Ignoring the
uncertainties in the limits on $k/f$, Perseus has the smallest range in $k/f$ as the two constraints come
from the weak shock ($k/f_{\rm shock}$) and the lack of strong shocks
in the rims ($k/f_{\rm sound}$).  
However the ranges in $k/f$ for the Northern ($390<k/f<570$) and
Southern ($760<k/f<1025$) bubbles do not
overlap, which is surprising.  However, if the uncertainties in the individual
limits are also included, then the ranges are much larger, and do
overlap ($200<k/f<1020$ and $390<k/f<1830$).  The values obtained are
in agreement with the range of $180<k/f<500$ obtained by
\citet{Celotti02} for the northern bubble.  This is assuming that the
synchrotron lifetime is less than the maximum age of the bubble, but there
is no physical requirement that this is the case.  All other clusters have an
upper limit on $k/f$ arising from the sound speed limit and two other
limits from physical arguments (buoyancy and refilling timescales).  

\begin{table*} \centering \caption{\scshape Physical $k/f$ Values}

\begin{tabular}{l l l l l l l l l l l l}
\hline
\hline
Cluster & Lobe$^{(1)}$ & Pressure  & $Re$ & Viscosity$^{(2)}$ & Energy$^{(3)}$  & $k/f_{\rm eq}$&
$k/f_{\rm sound}^{(4)}$  & $k/f_{\rm shock}$  & $k/f_{\rm buoyancy}$
&$k/f_{\rm refill}$\\
& & $(\rm eV/cm^3)$  & & $(10^{27}\cmsqps)$ & $(10^{58}\erg)$  & &   &   &  &\\

\hline
A133    & Rel, R& 52.8  & 4106 &  16.4  &       20.8   &   90.6  & $8.79\;^{424 }_{0.10}$   &   $-$&   $  57.3\;^{2760}_{0.68}$  &   $17.5\;^{843 }_{0.21}$\\
\\
A2052   & N, X  & 119   &  367 &  3.83  &       2.3    &   26.6  & $4.73\;^{25.7}_{0.81}$   &   $-$&   $  36.3\;^{197 }_{6.19}$  &   $12.6\;^{68.6}_{2.15}$\\
        & S, X  & 119   &  551 &  4.69  &       7.9    &   81.3  & $9.68\;^{52.6}_{1.65}$   &   $-$&   $  79.7\;^{433 }_{13.6}$  &   $24.0\;^{131 }_{4.10}$\\
\\
A2199   & E, R  & 116   &  754 &  3.77  &       4.2    &   21.5  & $3.85\;^{14.0}_{1.01}$   &   $-$&   $  7.90\;^{28.8}_{2.07}$  &   $3.02\;^{11.0}_{0.79}$\\
        & W, R  & 120   &  904 &  4.90  &       7.1    &   39.8  & $5.74\;^{21.0}_{1.50}$   &   $-$&   $  14.8\;^{53.9}_{3.87}$  &   $4.97\;^{18.1}_{1.30}$\\
        & I-E, R& 253   & 50.1 &  0.08  &       0.005  &    2.11 & $2.40\;^{12.1}_{0.46}$   &   $-$&   $  0.56\;^{2.84}_{0.11}$  &   $0.63\;^{3.17}_{0.12}$\\
        & I-W, R& 253   & 75.2 &  0.15  &       0.017  &    4.32 & $3.58\;^{18.0}_{0.68}$   &   $-$&   $  1.40\;^{7.06}_{0.274}$  &   $1.06\;^{5.32}_{0.20}$\\
\\
A4059   & N, R  & 123   &  148 &  0.69  &       0.065  &    9.97 & $4.15\;^{18.0}_{0.86}$   &   $-$&   $  14.9\;^{64.4}_{3.10}$  &   $8.85\;^{38.3}_{1.84}$\\
        &S, R   & 123   &  222 &  1.22  &       0.22   &    25.0 & $14.1\;^{61.1}_{2.94}$   &   $-$&   $  26.6\;^{115 }_{5.53}$  &   $21.5\;^{93.0}_{4.47}$\\
\\
Centaurus& E, R & 158   & 35.3 &  0.16  &       0.0045 &    1.57 & $1.21\;^{3.16}_{0.44}$   &   $-$&   $  2.18\;^{5.67}_{0.79}$  &   $0.86\;^{2.25}_{0.31}$\\
        &E, R   & 176   & 35.7 &  0.13  &       0.0037 &   2.00  & $1.26\;^{3.28}_{0.46}$   &   $-$&   $  2.03\;^{5.29}_{0.74}$  &   $0.88\;^{2.28}_{0.32}$\\
\\
Cygnus A& E, R  & 440   & 1552 &  6.21  &       116.0  &   35.5  & $1.50\;^{486 }_{0.37}$   &   $-$&   $  1.92\;^{6.23}_{0.48}$  &   $0.84\;^{2.72}_{0.21}$\\
        & W, R  & 484   & 2066 &  8.26  &       152.0  &   46.9  & $1.94\;^{6.27}_{0.48}$   &   $-$&   $  2.73\;^{8.84}_{0.68}$  &   $1.19\;^{3.86}_{0.30}$\\
\\
Hydra A & N, R  & 264   &  507 &  3.64  &       3.2    &   6.84  & $0.55\;^{3.11}_{0.07}$   &   $-$&   $  2.16\;^{12.2}_{0.29}$  &   $0.80\;^{4.53}_{0.11}$\\
        & S, R  & 238   &  416 &  2.16  &       1.5    &    4.45 & $0.47\;^{2.65}_{0.06}$   &   $-$&   $  0.91\;^{5.13}_{0.12}$  &   $0.50\;^{0.84}_{0.07}$\\
\\
M84     &N, R   & 57.2  & 71.8 &  0.52  &       0.019   &  205    & $201 \;^{207 }_{194 }$   &   $-$&   $  101 \;^{104 }_{98.1 }$  &   $268 \;^{276 }_{259 }$\\
        &S, R   & 57.2  & 73.5 &  0.52  &       0.020    &  196    & $110 \;^{114 }_{107 }$   &   $-$&   $  298 \;^{296 }_{279 }$  &   $202 \;^{208 }_{195 }$\\
\\
M87     & Bud, R& 660   & 56.2 &  0.22  &       0.029  &   56.2  & $23.9\;^{84.7}_{5.11}$   &  $-$ &   $  19.7\;^{70.0}_{4.22}$  &   $12.6\;^{44.7}_{2.70}$\\
\\
MKW3s   & S, R  & 28.2  & 1118 &  5.50  &       3.2    &    2.04 & $1.32\;^{1.66}_{1.05}$   &  $-$ &   $  0.78\;^{0.97}_{0.62}$  &   $0.76\;^{0.96}_{0.61}$\\
\\
Perseus & N, X  & 367   &  460 &  1.68  &       3.9    & 3966    & $ 573\;^{1020}_{292 }$   &  $390\;^{696 }_{199}$ & $  1031\;^{1839}_{526}$ & $278\;^{496}_{142}$\\
        & S, X  & 384   &  555 &  1.84  &       5.3    & 7262    & $1025\;^{1827}_{522 }$   &  $762\;^{1357}_{388}$ & $  1682\;^{2999}_{857}$&  $453\;^{807}_{231}$\\
\\
PKS1404 & E, X  & 115   & 53.5 &  0.47  &       0.009   &   503   & $ 343\;^{354 }_{332 }$   &  $-$  &   $  685\;^{707}_{662}$ &  $454\;^{468}_{438}$\\
        & W, X  & 118   & 53.7 &  0.55  &       0.011  &   327   & $ 327\;^{337 }_{316 }$   &  $-$  &   $  487\;^{502}_{470}$ &  $372\;^{384}_{359}$\\

\hline

\end{tabular} \label{kftable}
\begin{quote}
{\scshape Notes:} \\
(1) The codes for the Lobes are N---Northern, S---Southern, E---Eastern,
    W---Western, X---sizes from X-ray image, R---sizes from Radio image,
    I-E/W---Inner lobes in A2199, Bud---as described in
    \citet{FormanM87},  Rel---Relic source (for A133 as described in
    \citet{FujitaSarazin02}).\\

(2) The viscosity is estimated assuming that the flow is laminar and a Reynolds number of $1000$

(3) The energy quoted here is $E=PV$, so the values have to be
    multiplied by the appropriate $\gamma/(\gamma-1)$.\\

(4) The range on the limits on $k/f$ from the uncertainty in the spectral
    index are given by the maximum values (superscript) and minimum
    values (subscript).  The uncertainties from other parameters are
    shown in Fig. \ref{Clustererror}.\\

\end{quote}
\end{table*}
\begin{figure} \centering

\includegraphics[width=1.0 \columnwidth]{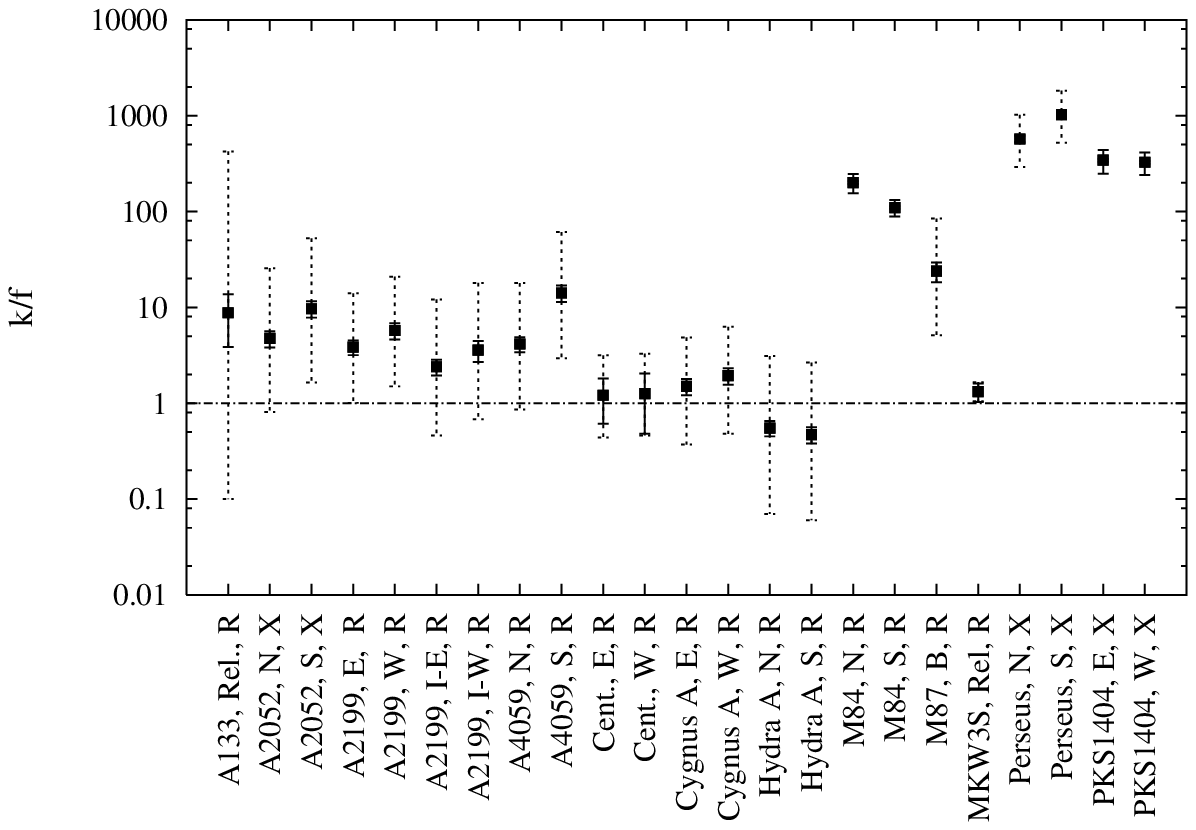}

\caption  {\label{Clustererror} \small{The value of $k/f$ calculated
    from the sound speed limit for each cluster
    analysed, along with the uncertainties arising from the
    uncertainties in $\alpha$ (the dotted bars) and from the
    uncertainties in the other physical parameters of the source (the
    solid bars).  The dotted line shows the minimum value of $k/f$
    possible from the assumptions used in the calculations.}}

\end{figure}

\begin{figure} \centering

\includegraphics[width=1.0 \columnwidth]{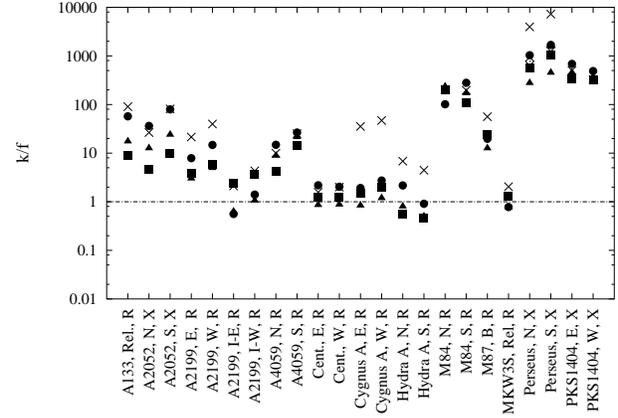}

\caption  {\label{Cluster} \small{The value of $k/f$ for each cluster
    analysed.  The $\blacksquare, \blacktriangle, \bullet$  and 
  $\lozenge$ symbols denote the $k/f$ values from the sound speed, shock,
  buoyancy and refilling timescales respectively.  The $\times$ symbol
  denotes the equipartition values}}

\end{figure}

The maximum and minimum values in the table above come from the
uncertainties in the values of the spectral index, $\alpha$, over the
radio lobe.  The resultant uncertainty in
$k/f$ from the uncertainty in $\alpha$ can be greater than an order of
magnitude as the spectral index appears as an exponent in Equation
\ref{Eparteqn}.  There are uncertainties in $k/f$ which come from the
uncertainties in the other observed values, however for most of the
clusters these are much smaller than the uncertainties from the those
in $\alpha$.  The limits on $k/f$
calculated from the lack of strong shocks in the rims (sound speed limit)
is plotted in Fig. \ref{Clustererror} with the uncertainties in the
limits on $k/f$ arising from the ranges in $\alpha$ and those from other
uncertainties plotted separately.  The dotted line
demarcates the minimum value possible for the assumptions used in the
calculations.  Each lobe is
plotted separately, identified by the same label as in Tables \ref{source prop}
and \ref{kftable}.  

The uncertainties in the spectral
index are generous --- $\alpha$ varies across the radio lobes, and so an
average value has been taken, with an uncertainty to cover the
range.  In some cases the values stated by different authors do
not match, and again an average has been taken, with the range
encompassing the values.

The variation in the limits on $k/f$ obtained from the
different timescales is shown in Fig. \ref{Cluster}.  On the whole,
the $k/f$ limiting values from different timescales fall within the
range of $k/f_{\rm sound}$ given by the uncertainties in $\alpha$.
It was not expected that the values would be exactly the same, but for
about half of the clusters the differences are less than $\sim \times
2.0$.  The remainders' are less than $\sim \times 8$ and these
disagreements arise primarily from differences in the timescales calculated (for
further discussion see Section \ref{timescales}).  As the values obtained for the
limits on $k/f$
from these different arguments are on the whole similar, it implies that
the actual limits on $k/f$ are around those values.  However all the
values have uncertainties of the same size as for the sound speed
(they have not been plotted for clarity), which reduces the
reliability of this conclusion.

For most of the bubbles, radio images at different
frequencies were obtained, and a separate limit on $k/f$ was calculated
for each.  The resultant values usually were all but identical, and
certainly agreed to within the uncertainties arising from all values
apart from the spectral index, however they have not been tabulated in this
work for brevity and clarity.  It was checked to see if the frequency
at which the flux was measured had some effect on the limit on $k/f$,
but none was found.  In the clusters where two bubbles are visible,
their $k/f$ limits broadly agree to within the uncertainties

\subsection{Timescales}\label{timescales}

The timescales (ages) for each lobe were calculated during the course
of this work.  The differences between our values and those presented
in the literature come mainly from the different sizes of
the bubbles used, and different cosmologies.  They have been tabulated
in the Appendix along with the
powers of the sources which have been calculated from these
timescales.  

For some of the bubbles the timescales obtained by the different
  estimates differ by up to a factor of 10.  These discrepancies are probably due to
the assumptions used in the calculation of the timescales.  

The sound speed timescale relies on the fact that the bubble is not
currently expanding supersonically and gives the time for the
expansion of the bubble.  From the observation that there are no strong
shocks at the rims we have assumed that the average expansion speed
was less than the current sound speed over the lifetime of the bubble.  
However, as we assume that the bubble is produced
by a relativistic jet and the presence of the shock in the X-ray
images of the Perseus cluster, the earlier expansion of the bubble must
have been supersonic, and so this lower limit on the timescale may be
an overestimate.  The lack of strong shocks in the rims still allows some
supersonic motion, \citet{Blanton01} state that $\mathcal{M}a<1.2$ for
A2052, so the sound speed may be an underestimate of the expansion speed.

The buoyancy timescale used is the time taken for a bubble of the size
observed to rise at its buoyancy velocity from the centre of the
gravitational potential to its current position, i.e. by its own
radius.  This is calculated assuming that the surrounding ICM is at
a constant pressure and density.  However as the bubble is expanding as it ages, the
initial stages are probably supersonic, the medium through which
it rises is likely to be non-uniform and in some cases the masses used
in the calculation are
estimates; the buoyancy velocity will change with the evolution of the
bubble, and so the timescale has uncertainties associated with it.

These two timescales are measuring the age of the bubbles
in two different ways, one the time for expansion to the current size at
the sound speed of the ICM at the edge of the bubble; the other the
time to travel to the current position at the buoyancy velocity.  This, along with the fact that calculation for the
expansion (sound speed) timescale uses twice the radius (see Section
\ref{Dataanal}) is the most likely cause of the discrepancies between
the timescales.  From the geometry of the bubbles, we have assumed
that the engine creating the bubble is not in the centre, but at one
side.  Therefore the part of the bubble which is furthest from the
source, the bubbles' diameter away, is the part whose speed must have been less than the sound
speed.  

In very young bubbles are the sound speed timescales less than the
buoyancy timescales e.g. M87 and A2199 (though these seem to be
expanding at a large distance from the central engine, and hence the
comparatively large buoyancy timescale).  The older, and usually larger,
bubbles have buoyancy timescales which vary from about equal (Cygnus
A) to ten times smaller than the sound speed timescale (A133). 

Both the buoyancy and refilling timescales depend upon the cluster
mass within the radius the bubbles are from the centre.  In the cases of A133 and A4059, the
masses have been estimated from a linear interpolation of the Abell
radius as given in \citet{Reiprich} and so the value obtained is
likely to be accurate only to factors of two to three.  In addition,
\citet{Slee01} also find that
the travel time of the relic in A133 is
longer than its age, a discrepancy which they propose can be resolved
by identifying the relic not with the cD galaxy, but the one 
labelled as G in their Figure 6, which gives a travel time of around
the age of the relic.  The relic does seem to be much larger than
others given its distance from the centre of the cluster, and so
projection effects may also play a part.  The timescales for A4059
also differ by factors of up to three, and if the mass estimate is
reduced from $1.5/1.1 \times 10^{14} {\rm M_{\odot}}$ for the
Northern and Southern bubbles respectively to $\sim 0.20 \times
10^{14} {\rm M_{\odot}}$, then the timescales become more comparable
($\times 3$ compared to $\times 6$ different).  This
is a reduction of $\sim 90\%$, however
a linear interpolation of the mass from the Abell radius 
overestimates the masses of Perseus and A2199 (inner bubbles) by more
than a factor of ten. These new masses for A4059 give $t_{\rm buoy} = 0.71 \times 10^7 \yr$,
$t_{\rm ref} = 2.0 \times 10^7 \yr$,
$k/f_{\rm buoy} = 8.9$ and $k/f_{\rm ref} = 3.51$ for the Northern
  bubble and $t_{\rm buoy} = 0.43 \times 10^7 \yr$, $t_{\rm ref} = 1.7 \times 10^7 \yr$,
$k/f_{\rm buoy} = 31.9$ and $k/f_{\rm ref} = 9.92$ for Southern
bubble.  The $k/f$ values do not change radically ($\sim 50 \%$ drop) as a result of
this.


The form for the synchrotron lifetime (Equation \ref{synctime}) is
only valid  for sufficiently large magnetic fields.  Inverse Compton (IC)
losses dominate when the energy density of the Cosmic Microwave
Background is equal to that of the magnetic field, $U_{\rm
  B}=B^2/8\pi$, which corresponds to
$B=B_{\rm CMB}$.  Although all the clusters analysed are in the local
Universe, it was checked whether the inferred magnetic field was
larger than $B_{\rm CMB}$.  The lifetime determined
from this limiting magnetic field for $1\ghz$ electrons ($t_{\rm CMB}$) was also calculated and is tabulated
in the Appendix.  Only A133 has a timescale, inferred from the sound
speed, that is larger than $t_{\rm CMB}$.  The magnetic field is also
smaller than $B_{\rm CMB}$, and so IC losses are important in this
cluster.  The Eastern lobe of Cygnus A has a
magnetic field from the refilling timescales that is less than 10\%
larger, and so IC losses may be important in this cluster as well.

The sound speed timescale gives a lower limit on the
age, and hence an upper limit on the magnetic field, and so the IC
effects are likely to be important for more clusters than just the two
mentioned above, as the actual age of the bubble is probably larger,
and so the magnetic field smaller.  

\section{Discussion}\label{discussion}

Initially it had been hoped that limits on $k/f$ for the clusters were all
of the same order of magnitude, however, from Fig. \ref{Clustererror} it is
apparent that they are not.  The minimum possible value for $k/f=1$ is
shown by the dotted line.  Hydra A
is the only source whose $k/f$ value falls below this line, though the
uncertainties are such that it could have a value greater than one.  A
possible reason for this source having such a low $k/f$ value is that
the bubbles may have punched their way through the ICM.  Therefore the
bubbles are no longer being contained and so are no longer in
the type of equilibrium that has been assumed in the calculations.

The calculations have shown that all the lobes cannot be in
equipartition. Although some of the limits on $k/f$ are larger than
the equipartition $k/f$ values, this is due to the form of the
dependence of $k/f$ on the magnetic field (see Fig. 7 of \citet{Celotti02}).  All the timescales derived
from observational data indicate that the bubbles cannot be in
equipartition.  The only timescales that
allow equipartition are the minimum and maximum from the dynamical
constraints (Section \ref{DynCons}), and only for some clusters. 

\begin{figure} \centering

\includegraphics[width=1.0 \columnwidth]{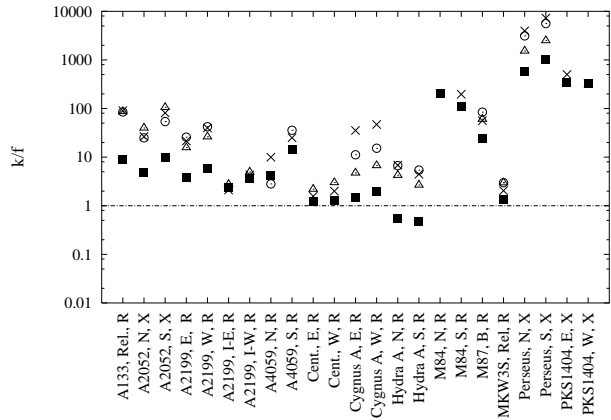}

\caption  {\label{LimMax} \small{The value of $k/f$ for each cluster
    analysed.  The $\blacksquare, \circ$  and $ \triangle$ symbols
  denote the $k/f$ values from the sound speed, minimum and maximum
  timescales respectively (Section \ref{Dynconst}).  The $\times$
  symbol denotes
  the equipartition values.}}

\end{figure}

The limits on $k/f$ from the maximum and minimum ages of the hole as
calculated from the dynamical constraints arguments in Section
\ref{Dataanal} are shown in Fig.
\ref{LimMax}, along with the equipartition values and the sound speed limits
for comparison.
These limits are higher than
the ones obtained from the sound speed for many clusters.  This implies that the
assumptions used in the calculation of these limits on $k/f$ may not be
valid.  Another indication that this may be so is the fact that the
timescales are inconsistent with each other --- the maximum age of the
hole is frequently less than the minimum age of the hole (see the Appendix
for the values), and, on the whole, both are less than the sound speed
values.  This can be seen in Fig. \ref{Dynconst}  where the maximum
allowed age for the northern bubble in the Perseus cluster is less than
$10^7$ years, whereas the values calculated from the sound speed and
shock front arguments give $2.2\times 10^7$ and $3.2\times10^7$ years
respectively.  We therefore believe that the assumptions made in
the derivation and calculation of these values are not valid for these
bubbles.  The largest assumption are that the bubbles
have been expanding sub-sonically and in a medium of constant pressure,
which is unlikely, as there is a shock front
visible in the Perseus X-ray images an the temperature and density of
the ICM varies with distance from the centre of the cluster.  Another assumption
used is that they have not buoyantly detached from the source, which
in some cases must have occurred (e.g. A133 and MKW3s).  

\begin{figure} \centering

\includegraphics[width=1.0 \columnwidth]{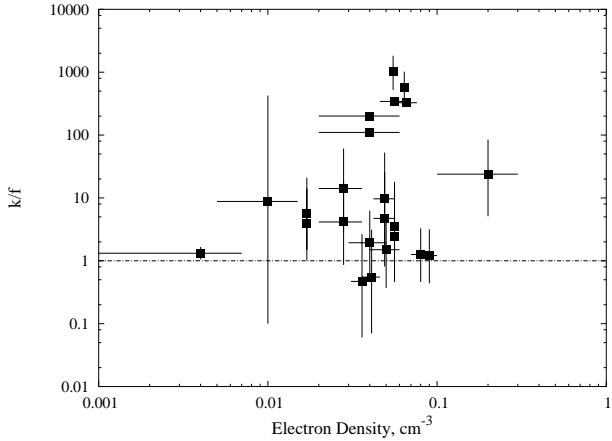}

\caption  {\label{n_e} \small{The distribution of $k/f$ versus electron
  density.  The errors in $k/f$ come from the uncertainty in $\alpha$.}}

\end{figure}

\begin{figure} \centering

\includegraphics[width=1.0 \columnwidth]{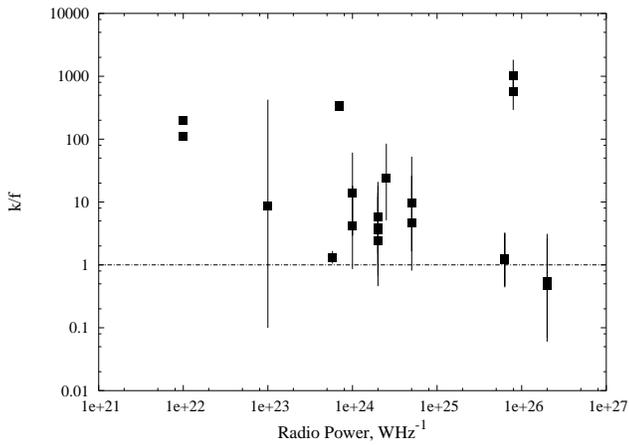}

\caption  {\label{radpower} \small{The distribution of $k/f$ versus
    the radio power of the source at 5 GHz.  The errors in $k/f$ come from the uncertainty in $\alpha$.}}

\end{figure}

\begin{figure} \centering

\includegraphics[width=1.0 \columnwidth]{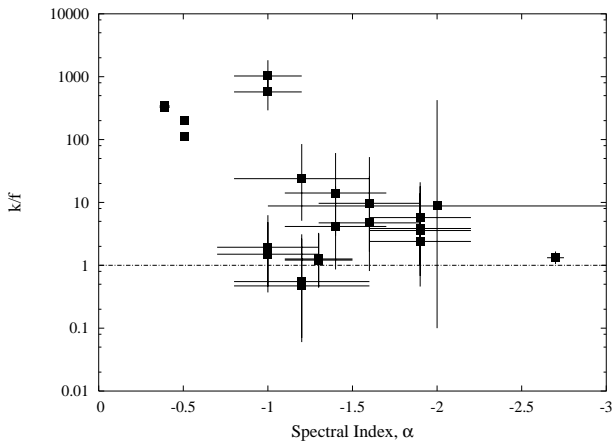}

\caption  {\label{Alpha} \small{The distribution of $k/f$ versus
    $\alpha$.  The errors in $k/f$ come from the uncertainty in $\alpha$.}}

\end{figure}

Although the $k/f$ values calculated are upper limits, treating them
as absolute values, it was
investigated if there was any distribution of $k/f$ with the
physical parameters of the sources.  The only parameters which came up
with possible trends were the electron density, the 5 GHz radio power of
the source and the spectral index of the lobes (Figs. \ref{n_e},
\ref{radpower} and \ref{Alpha}).  In the plot of $k/f$ against
electron density there seems to be a cut-off line
from the lower left to the upper right-hand side of the scatter plot,
with all the clusters falling to the right of this line.  There seems
to be a trend with $5\ghz$
radio power, with low power sources having a high $k/f$, and high
power sources having a low $k/f$.  However the Perseus cluster is an
exception to this last trend (the
points with a $k/f\sim 500$ at a radio power of $\sim 3 \times
10^{25}{\rm\thinspace W~Hz^{-1}}$).  Any effect from beaming has not been taken
into account, but as the bubbles seem to be moving
across the line of sight, it is unlikely that there are large beaming
effects.  A similar trend may be present in the plot of
$k/f$ versus $\alpha$.  It is also obvious in
this plot of the large
effect an uncertainty in the spectral index has on the uncertainty in
$k/f$.  In the above discussion the uncertainties on $k/f$ values have
not been taken into account, however due to the size of the
uncertainties any firm conclusions being drawn from the plots.  We
believe that there is no clear correlation of $k/f$ with any physical parameter of
the bubble.


\begin{figure} \centering

\includegraphics[width=1.0 \columnwidth]{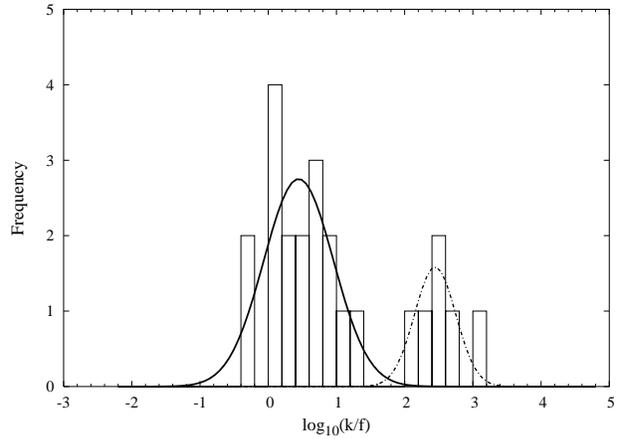}

\caption  {\label{Bimodal} \small{The distribution of the numbers of
    bubbles with given ${\rm log}_{10}(k/f_{\rm sound})$ values along with the
    best fitting Gaussian distributions from least-squares analysis.}}

\end{figure}

\begin{figure} \centering

\includegraphics[width=1.0 \columnwidth]{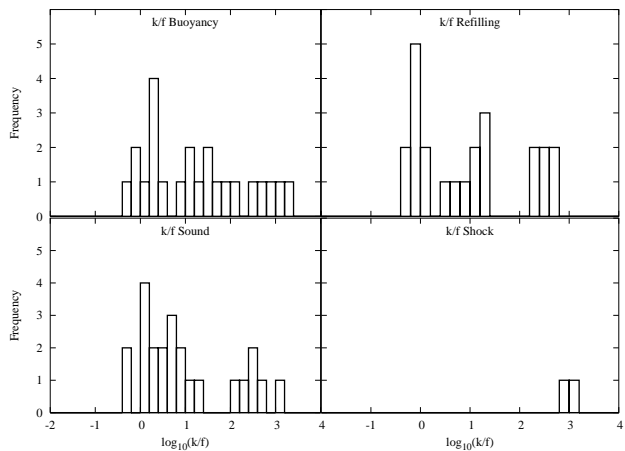}

\caption  {\label{Multiplot} \small{The distribution of the numbers of
    bubbles with given ${\rm log}_{10}(k/f)$ values for the sound
    speed, buoyancy, refilling and shock timescale calculations.}}

\end{figure}

As there was no firm trend of $k/f$ with any of the physical
parameters of the lobes or the source, other possible distributions
were looked for, and it was noticed that the $k/f_{\rm sound}$ values
were clustered - the value for some clusters occurred around 3 and others occurred around 300.
The number of bubbles with a ${\rm log}_{10}(k/f_{\rm sound})$ values in a given range
were binned (Fig. \ref{Bimodal}).  The distribution appears to be
bimodal, with a sample size
for the distribution of 23.  When the lower populations' values are at
the maximum allowed by their uncertainty and the higher
populations' values at their minimum, the two populations
overlap.  Therefore there may be a
continuous distribution of $k/f_{\rm sound}$ which we have not seen as our sample
size is small. However,
with the uncertainties that we have quoted on $k/f$ then no
conclusions can be definite.  
The distribution of $k/f$ values from all timescales was tested using the
\emph{EMMIX}\footnote{http://www.maths.uq.edu.au/$\sim$gjm/emmix/emmix.html}
program, a more recent
version of the {\sc KMM} algorithm described in \citet{AshmanBirdZepf},
which uses expectation maximisation to perform a maximum likelihood calculation.
The values from the
sound speed argument are most likely to be described by a bimodal
Gaussian distribution, however the buoyancy values are best
fitted by a trimodal distribution, and the
refilling values are all but equally likely to be fitted by two,
three and four populations.  However, all are unlikely to be
described by a single Gaussian in log space. The binned $k/f$ values for all of the
physical timescale calculations
are shown in Fig. \ref{Multiplot} for comparison.  


The data from the sound speed were then fitted using least squares by two Gaussians, which
are shown in the Fig. \ref{Bimodal}, with means of
${\rm log}_{10}(k/f_{\rm sound})=0.44\pm0.09$  and $2.45 \pm 0.06$ ($\sim2.3$ and
$\sim280$).  The clustering analysis produced the following means
${\rm log}_{10}(k/f_{\rm sound})=0.44$ and $=2.51$ ($\sim2.8$ and
$\sim320$), which are within the range from the
uncertainties given by least squares fitting.
  

\begin{table*} \centering \caption{\scshape \label{ghostlobes}: Lower
    limits on $k/f$ from Ghost Bubbles in Perseus}
\begin{tabular} {l |l l l l |l l l l}
\hline
\hline
Lobe$^{1}$ & \multicolumn{4}{c} {$k/f$} & \multicolumn{4}{c} {Timescales, $10^7 {\rm yr}$} \\
& $k/f_{\rm eq}$&
$k/f_{\rm sound}$  & $k/f_{\rm buoyancy}$
& $k/f_{\rm refill}$&$t_{\rm CMB}^{(2)} $& $t_{\rm sound}$& $t_{\rm buoy}$ & $t_{\rm refill}$\\
\hline

West & 4271 & $2267\;^{8691}_{526}$ & $252\;^{968}_{58.6}$ & $350\;^{1343}_{81.3}$ & 13.9 & 0.795 & 7.53 & 5.42 \\
South & 4460 & $1337\;^{5127}_{310}$ & $304\;^{1165}_{70.5}$ & $346\;^{1327}_{80.3}$ & 13.9 & 1.59  & 7.15 & 6.27 \\
\hline

North& 3996 & $573\;^{1020}_{292}$   &  $  1031\;^{1839}_{526}$  & $278\;^{496}_{142}$ &  13.9 & 2.18  & 1.19 & 4.50 \\
South& 7262 & $1025\;^{1827}_{522}$  &  $  1682\;^{2999}_{857}$  &  $453\;^{807}_{231}$&  13.9 & 2.15  & 1.30 & 4.90 \\

\hline
\end{tabular}
\begin{quote}
{\scshape Notes:}\\
(1) The lower limits on $k/f$ for Ghost Bubbles in the Perseus Cluster
(West and South).  The inner bubbles' values for the upper limits on $k/f$ are shown for
comparison (North and South).

(2) The timescale for the bubble calculated from the magnetic field
    which produces the same energy density as that of the CMB at the
    redshift of the cluster for electrons radiating at $1\ghz$
\end{quote}
\end{table*}
 
In order to allow the discussion of a bimodal population of
$k/f$ it was investigated whether it was possible to obtain lower
limits on $k/f$ for the upper population.  If a lower
limit could be obtained and is larger than the upper limits of the
lower population then the bimodality of the distribution of $k/f$
can be discussed with more reliability.

A lower limit was obtained using the ghost bubbles in Perseus, the
only cluster where this is possible to some degree of accuracy.  A
calculation similar to that for the upper limits was used and the
results are shown in Table \ref{ghostlobes}.  As there is no $\ghz$
radio emission from the ghost bubbles, the electrons have all aged.  Therefore the age of the bubble must be
\emph{longer} than synchrotron cooling time
of the $\ghz$ electrons. Radio
emission at $330 \mhz$ was analysed to obtain the energy in electrons
radiating radiating between $\nu_1 = 10\mhz$ and $\nu_2 = 10\ghz$ (see
Equation \ref{Eparteqn}) using a steep spectral index of $\alpha =
-2.0 \pm 0.5$..  

The calculation assumes that the electron
distribution is still described by a single spectral index over the
above range of frequencies.  These lower limits on $k/f$ are above
most of the upper limits on $k/f$ obtained for the lower population in
Figure \ref{Bimodal} (ignoring errors on the limits).
There are again discrepancies in the timescales, mainly due to the
fact that the sound speed timescale measures the expansion time of the
bubble while the buoyancy timescale measures the travel
time to the bubbles' current location.  In some
cases the lower limit obtained is larger than the upper limits for the
inner lobes.  However, if the ranges in the limits are included then
there is overlap between the limits.  Also, it is reasonable to use
the limit on $k/f$ calculated from
the buoyancy timescale rather than the sound speed timescale as the
lower limit, as these bubbles have detached from
the centre and so are rising buoyantly through the ICM.  Using this
timescale, the lower limits on $k/f$ are lower than the upper limits from
the younger bubbles. 

From the observation that there is smooth spectral ageing across the
inner lobes \citep{Celotti02} and no $\ghz$ emission from the outer ghost bubbles we are
able, in the special case of the Perseus cluster, to obtain lower as
well as upper limits on the value for $k/f$.  It is
constrained to be between $\sim 100$ and $\sim 1000$ with the assumption
that there is no large change in resulting from the bubbles' detaching
from the central engine or its age.  The upper limits on
$k/f$ for most of the other clusters in the sample are not compatible
with this range, which implies that there is at least a large spread
in the values for $k/f$, if not a bimodal distribution.  Although the
sample size is small (23) we now go on to discuss the implications of a
bimodal distribution.

\section{Interpretation}\label{implications}

To obtain a high value of $k/f$ then either $k$ is intrinsically high
or $f \ll 1$ .  If $f\sim 1$, then $k$ is high and the
particles corresponding to the observed synchrotron radio
emission do not account for all the particles required for there to be
pressure equilibrium.  Hence, for M84, Perseus and
PKS 1404-267, all of which have high values for $k/f$, the number of
particles emitting synchrotron radiation are not sufficient for the bubble
to be in pressure equilibrium.  These bubbles, of course, may not be
in pressure equilibrium, however the lack of particles would imply that
the bubbles are collapsing, which assuming that they are still
``powered,'' is unlikely.  The extra particles required could come from
various sources -- particles radiating at energies less than $10\mhz$,
any non-relativistic (thermal) component which has been swept up into
the bubbles during their creation and also turbulence.  If the jet creating
the bubbles were electron-proton, then the protons would be
undetectable from the radio emission alone, however they would exert a
pressure on the surrounding ICM, and so could
account for the extra pressure in the bubble.  This implies that there
are two types of jets producing radio bubbles --- heavy, electron-proton
jets giving a high value for $k/f$, and light, electron-positron jets
giving a low value for $k/f$.   Both
types of jets have been discussed in the past --- \citet{CelottiACFjets} present
arguments based on the physical properties of parsec scale radio
sources that radio jets are electron-proton; whereas \citet{ReynoldsACFjets}
argue that, at least for M87, the jets are electron-positron.  

Another effect may be due to the spectral ageing of the electrons in
the radio plasma.  Once the bubble plasma is no longer ``powered'' then the
electrons will radiate away their energy, with the highest energy ones
losing their energy quickest.  Assuming an initial power law spectrum,
the spectrum will steepen at
the high frequency end.  If the spectral index that has been used in
the calculations comes from the part of the spectrum that has steepened, as it is
extrapolated back down to $10\mhz$, the value for the flux emitted at
lower energies will be an overestimate.  Therefore the number of
lower-energy particles will also be overestimated, and so pressure
equilibrium would seem to be achieved with the ``observed'' particles,
so giving a low $k$ value, even though these particles do not exist
and $k$ is high.  This would imply that the low
values for $k/f$ obtained for some clusters are because a spectral index that is
too steep has been used in the calculation.  MKW3s and A133 could be
such sources.  Their bubbles appear to have detached, they have very steep spectral indices and their $k/f$ values
are consistent with one.  However the radio emission from other bubbles do not have
unusually steep spectra and hence it is unlikely that this situation
occurs in the majority of the bubbles.

If $f$ is allowed to
vary, then it would be possible to obtain a high $k/f$ with a low $k$,
so the particle energy factors are the same for all clusters, and all
that varies is the volume filling factor.  Using the means of the
two Gaussian distributions from Section \ref{discussion}, a  volume
filling fraction of $\sim0.01$ for the clusters with a higher $k/f$ value
would mean that their $k$ values were around $2-3$, similar to the
other sources.  However this begs the question - why do there seem
to be two types of bubbles - ones with $f\sim1$ and ones with
$f\sim0.01$?  A bimodal distribution seems unlikely in this case, and
therefore if the variation in $k/f$ is due to variation on $f$ then
there is probably a continuum, possibly shown by $k/f_{\rm buoy}$, and
the clusters that we have chosen appear to give a bimodal distribution of $k/f_{\rm sound}$.  
If the distribution of the radio-emitting plasma were filamentary, then
$f$ may have a value $\sim0.01$, and a number of radio lobes do show
filamentary structure, e.g. Fornax A
\citep{FomalontFornax}, M87 \citep{Owen00M87} and Cygnus A
\citep{PerleyCygnusA}.  

In Section \ref{nonuniform} we have calculated the effect of a
simple non-uniform magnetic field, which approximately doubled
the value of $k/f$ over that obtained from a uniform field.  Hence if the magnetic field were highly
non-uniform then the $k/f$ value may increase a substantial amount.
As the radio emission is filamentary in some of the lobes which have
been analysed, the magnetic field may be as well.

A combination of the two effects may also fit the data.
\citet{ACF_Halpha_PER03} showed that the ghost bubbles in Perseus drag
up thermal gas (H$\alpha$) as they rise through the ICM.  If this gas
is also entrained and mixed with the relativistic plasma during the
creation of the bubbles, then the volume filling
factor of the radio plasma would be less than one, and, there are extra
particles present, meaning that $k$ would be greater than one.  This
would also imply that there is some evolution of $k/f$ with time,
however plotting the $k/f$ value against the corresponding timescale
shows no clear correlation.  However \citet{Schmidt2002} state that the lobes are
essentially devoid of thermal gas, though if it were hotter than $\sim
11\kev$ it would not have been detected. 

The r\^ole of $f$ could also be an explanation, i.e. whether $f$
represents the fraction of the volume of the bubble which is occupied
by radio plasma, and the remainder is thermal gas, or whether $f$ is
the fraction of the volume filled with detectable radio emitting
plasma, and the remainder in this case is filled with relativistic
plasma where the electrons (and positrons) have aged, but the magnetic
field (and protons) have not lost any of their energy.  For the
sources with a low $k/f$ and where the radio emission does not fully
occupy the X-ray bubble, e.g.A2052, rather than using the flux from
the total observes radio emission, the flux from the lower surface
brightness region were used.  Interpolating this over the whole volume
would mean that fewer particles would be inferred from the radio
emission, and so the value for $k/f$ would increase.

\subsection{Re-acceleration} \label{Reacc}

For there to be re-acceleration in the bubble the magnetic field has to
be sufficiently strong.  In this case the synchrotron cooling time of
the electrons could be very short and we would not expect to detect
GHz emission from the bubble.  Nevertheless, if there is
re-acceleration then the electrons may still be radiating, from this
we would infer a long cooling time.  Combining Equations \ref{k/f} and
\ref{synctime} in the regime where the $B^2$ term is small enough to be ignored,
then $k/f \propto t_{synch}^{-1}$, and so an inferred cooling time
that is long would lead to a limit on $k/f$ that is lower than appropriate.

This could explain the shape and spread of the distribution of the
limits on $k/f$.  If the bubbles which have a limit on $k/f$ which is
small have ongoing re-acceleration of electrons, then even though the
cooling time is short, they have detectable radio emission.  Hence we
have assumed that the cooling time is long, of order the age of the
bubble, and so the calculated limit on $k/f$ is smaller than if the
actual cooling time had been used.  In this case the bubbles which
have a limit on $k/f$ which is large have comparatively little
re-acceleration, and the limits on $k/f$ obtained are close to those which have
been used in the literature.

The limits on $k/f$ cannot be raised indefinitely - there is a maximum
value for the limit (Section \ref{Dataanal}).  However if all the bubbles
whose limit on $k/f$ is in the lower population of the bimodal
distribution arising from the lack of strong shocks in the rims have
their limit on $k/f$ raised to the maximum possible, whereas the others
remain where they are; then a multimodal distribution is no more
likely that a single Gaussian.  The increase in the magnetic field
required to raise the limits on $k/f$ to their maximum range from  1.5
(A2199, I-E) to 15 (Cygnus A, E) times that of the one inferred
from the synchrotron cooling time with no re-acceleration.  These
fields lead to cooling times for 1 GHz electrons that are 0.54 to 0.017 times those
inferred from the age of the bubble.   Hence many cycles of
re-acceleration are required such that the electrons are detected,
however $\sim 60$ cycles seems unlikely.

There are two possibilities for how re-acceleration occurs - either
there is a population of electrons whose number is almost fixed, and
they are re-accelerated repeatedly many times over the age of the
bubble so that they are still radiating today.  As the radio
luminosity $L_{\rm radio} \propto N_0 B^{1-\alpha}\nu^{\alpha}{\rm
  dV}$, then for a given luminosity, if the magnetic field is
stronger, then there are fewer electrons present.  If the
electrons are re-accelerated, then it is likely that other particles
present (e.g.\ protons \& ions) will also be re-accelerated. This
would have a minimal effect on $k/f$ as, although there are fewer
particles present, they have a higher energy from the
re-acceleration.  These two effects cancel each other out to some
extent and so $k$ is about the same.  There are no new extra particles
present in the bubble in this case, and so $f$ of the synchrotron
emitting plasma is also about the same, hence there is little change
in $k/f$ and so this is unlikely as an explanation.

In the other case the electrons are
re-accelerated once, in a hotspot say, and then flow throughout the
bubble where they age rapidly, but are continuously replaced and so
there is still radio emission present today. Cygnus A has a radio
structure which seems to match the latter explanation -
i.e. the strongest radio emission is at the hotspot, and then the
image appears to show that the plasma flows away and becomes less
radio bright.  In this explanation there would be a population
of aged electrons which reduces the volume filling fraction of the
emitting relativistic plasma and increases $k$ as the aged electrons
also exert a pressure.  In this case we only detect $\sim
t_{\rm sync}/t_{\rm bubble}$ of the particles present.  The problem with this
model is that for the electrons to flow throughout the bubble before
they age they have to travel at $\sim 0.3{\rm\thinspace c}$ (for a cooling time
of $\sim 10^5\yr$ and a bubble radius of $5\kpc$).


\section{Conclusions}

From a sample of low redshift clusters with clear radio bubbles which
are coincident with decrements in the X-ray emission, we have
determined limits on  $k/f$, where $k$ is the
ratio of the total relativistic particle energy to that in electrons
radiating between $10\mhz$ to $10\ghz$ and $f$ is the volume
filling factor of the relativistic plasma.  We find that no bubble has simple
equipartition between the pressures from the relativistic
particles and the magnetic field.  $k/f$ was found to have
no strong dependence on any physical parameter of the host cluster,
however there seemed to be two populations.  One set of clusters had
a $k/f$ value around 2, the remainder had a value of around 300.  This
apparent bimodality of the distribution of $k/f$ could be explained in
various ways.  The jets creating the bubbles are of two types ---
electron-positron, which would have a low value for $k$ (for $f \sim
1$), and electron-proton, where the protons are the extra particles
required to maintain pressure equilibrium, but as they are unseen in
the radio emission $k$ is high.  Spectral ageing of the radio plasma
steepens the spectrum, so giving the impression that there are more
particles present.  A bimodality in the
volume filling fraction which could be caused by either a non-uniform
magnetic field, or a filamentary structure in the lobes. If thermal
plasma is entrained during the formation of the bubbles this would reduce the
volume filling factor and provide extra particles, resulting in the
calculated values.  Variations in the amounts of re-acceleration may
also produce the observed distribution.

\section*{Acknowledgments}

We thank Greg Taylor for valuable help with obtaining some of the
radio images, Annalisa Celotti for helpful discussions, Guy Pooley and Rosie Bolton for guidance with AIPS,
and the referee, Eugene Churazov, for many helpful comments and suggestions.
ACF and RJHD acknowledge support from The Royal Society
and PPARC respectively.

\bibliographystyle{mnras} 
\bibliography{mn-jour,dunn}

\begin{thebibliography}{}

\bibitem[\protect\citeauthoryear{{Alexander}, {Brown}, \& {Scott}}{{Alexander}
  et~al.}{1984}]{Alexander84}
{Alexander} P., {Brown} M.~T.,  {Scott} P.~F., 1984, \mnras, 209, 851

\bibitem[\protect\citeauthoryear{{Andernach} et~al.}{{Andernach}
  et~al.}{1988}]{Andernach}
{Andernach} H., {Han Tie} , {Sievers} A., {Reuter} H.-P., {Junkes} N.,
  {Wielebinski} R., 1988, \aaps, 73, 265

\bibitem[\protect\citeauthoryear{{Ashman}, {Bird}, \& {Zepf}}{{Ashman}
  et~al.}{1994}]{AshmanBirdZepf}
{Ashman} K.~A., {Bird} C.~M.,  {Zepf} S.~E., 1994, \aj, 108, 2348

\bibitem[\protect\citeauthoryear{{B{\^ i}rzan} et~al.}{{B{\^ i}rzan}
  et~al.}{2004}]{Birzan04}
{B{\^ i}rzan} L., {Rafferty} D.~A., {McNamara} B.~R., {Wise} M.~W.,  {Nulsen}
  P.~E.~J., 2004, \apj, 607, 800

\bibitem[\protect\citeauthoryear{{Bettoni} et~al.}{{Bettoni}
  et~al.}{2003}]{Bettoni03}
{Bettoni} D., {Falomo} R., {Fasano} G.,  {Govoni} F., 2003, \aap, 399, 869

\bibitem[\protect\citeauthoryear{{Blanton}, {Sarazin}, \& {McNamara}}{{Blanton}
  et~al.}{2003}]{Blanton03}
{Blanton} E.~L., {Sarazin} C.~L.,  {McNamara} B.~R., 2003, \apj, 585, 227

\bibitem[\protect\citeauthoryear{{Blanton} et~al.}{{Blanton}
  et~al.}{2001}]{Blanton01}
{Blanton} E.~L., {Sarazin} C.~L., {McNamara} B.~R.,  {Wise} M.~W., 2001, \apjl,
  558, L15

\bibitem[\protect\citeauthoryear{{B{\"o}hringer} et~al.}{{B{\"o}hringer}
  et~al.}{1993}]{Bohringer}
{B{\"o}hringer} H., {Voges} W., {Fabian} A.~C., {Edge} A.~C.,  {Neumann} D.~M.,
  1993, \mnras, 264, L25

\bibitem[\protect\citeauthoryear{{Brown} \& {Burns}}{{Brown} \&
  {Burns}}{1991}]{BrownBurns91}
{Brown} D.~L.,  {Burns} J.~O., 1991, \aj, 102, 1917

\bibitem[\protect\citeauthoryear{{Burbidge}}{{Burbidge}}{1959}]{Burbidge}
{Burbidge} G.~R., 1959, \apj, 129, 849

\bibitem[\protect\citeauthoryear{{Burns}}{{Burns}}{1990}]{Burns90}
{Burns} J.~O., 1990, \aj, 99, 14

\bibitem[\protect\citeauthoryear{{Burns}, {Schwendeman}, \& {White}}{{Burns}
  et~al.}{1983}]{Burns83}
{Burns} J.~O., {Schwendeman} E.,  {White} R.~A., 1983, \apj, 271, 575

\bibitem[\protect\citeauthoryear{{Cao} \& {Rawlings}}{{Cao} \&
  {Rawlings}}{2004}]{CaoRawlings04}
{Cao} X.,  {Rawlings} S., 2004, \mnras, 349, 1419

\bibitem[\protect\citeauthoryear{{Carilli} et~al.}{{Carilli}
  et~al.}{1991}]{Carilli91}
{Carilli} C.~L., {Perley} R.~A., {Dreher} J.~W.,  {Leahy} J.~P., 1991, \apj,
  383, 554

\bibitem[\protect\citeauthoryear{{Celotti} \& {Fabian}}{{Celotti} \&
  {Fabian}}{1993}]{CelottiACFjets}
{Celotti} A.,  {Fabian} A.~C., 1993, \mnras, 264, 228

\bibitem[\protect\citeauthoryear{{Choi} et~al.}{{Choi}
  et~al.}{2004}]{ChoiA4059}
{Choi} Y., {Reynolds} C.~S., {Heinz} S., {Rosenberg} J.~L., {Perlman} E.~S.,
  {Yang} J., 2004, \apj, 606, 185

\bibitem[\protect\citeauthoryear{{Churazov} et~al.}{{Churazov}
  et~al.}{2001}]{Churazov01}
{Churazov} E., {Br{\" u}ggen} M., {Kaiser} C.~R., {B{\" o}hringer} H.,
  {Forman} W., 2001, \apj, 554, 261

\bibitem[\protect\citeauthoryear{{Churazov} et~al.}{{Churazov}
  et~al.}{2000}]{Churazov00}
{Churazov} E., {Forman} W., {Jones} C.,  {B{\" o}hringer} H., 2000, \aap, 356,
  788

\bibitem[\protect\citeauthoryear{{Drinkwater} et~al.}{{Drinkwater}
  et~al.}{1997}]{Drinkwater}
{Drinkwater} M.~J. et~al., 1997, \mnras, 284, 85

\bibitem[\protect\citeauthoryear{{Fabian} et~al.}{{Fabian}
  et~al.}{2002}]{Celotti02}
{Fabian} A.~C., {Celotti} A., {Blundell} K.~M., {Kassim} N.~E.,  {Perley}
  R.~A., 2002, \mnras, 331, 369

\bibitem[\protect\citeauthoryear{{Fabian} et~al.}{{Fabian}
  et~al.}{2003a}]{ACF_deep_PER03}
{Fabian} A.~C., {Sanders} J.~S., {Allen} S.~W., {Crawford} C.~S., {Iwasawa} K.,
  {Johnstone} R.~M., {Schmidt} R.~W.,  {Taylor} G.~B., 2003a, \mnras, 344, L43

\bibitem[\protect\citeauthoryear{{Fabian} et~al.}{{Fabian}
  et~al.}{2003b}]{ACF_Halpha_PER03}
{Fabian} A.~C., {Sanders} J.~S., {Crawford} C.~S., {Conselice} C.~J.,
  {Gallagher} J.~S.,  {Wyse} R.~F.~G., 2003b, \mnras, 344, L48

\bibitem[\protect\citeauthoryear{{Fabian} et~al.}{{Fabian}
  et~al.}{2000}]{ACF_complex_PER00}
{Fabian} A.~C. et~al., 2000, \mnras, 318, L65

\bibitem[\protect\citeauthoryear{{Finoguenov} \& {Jones}}{{Finoguenov} \&
  {Jones}}{2002}]{FinoguenovJones02}
{Finoguenov} A.,  {Jones} C., 2002, \apj, 574, 754

\bibitem[\protect\citeauthoryear{{Fomalont} et~al.}{{Fomalont}
  et~al.}{1989}]{FomalontFornax}
{Fomalont} E.~B., {Ebneter} K.~A., {van Breugel} W.~J.~M.,  {Ekers} R.~D.,
  1989, \apjl, 346, L17

\bibitem[\protect\citeauthoryear{{Forman} et~al.}{{Forman}
  et~al.}{2003}]{FormanM87}
{Forman} W. et~al., 2003, astro-ph/0312576

\bibitem[\protect\citeauthoryear{{Fujita} et~al.}{{Fujita}
  et~al.}{2002}]{FujitaSarazin02}
{Fujita} Y., {Sarazin} C.~L., {Kempner} J.~C., {Rudnick} L., {Slee} O.~B.,
  {Roy} A.~L., {Andernach} H.,  {Ehle} M., 2002, \apj, 575, 764

\bibitem[\protect\citeauthoryear{{Johnstone} et~al.}{{Johnstone}
  et~al.}{2002}]{JohnstoneA2199}
{Johnstone} R.~M., {Allen} S.~W., {Fabian} A.~C.,  {Sanders} J.~S., 2002,
  \mnras, 336, 299

\bibitem[\protect\citeauthoryear{{Johnstone}, {Fabian}, \&
  {Taylor}}{{Johnstone} et~al.}{1998}]{JohnstonePKS}
{Johnstone} R.~M., {Fabian} A.~C.,  {Taylor} G.~B., 1998, \mnras, 298, 854

\bibitem[\protect\citeauthoryear{{Komissarov} \& {Gubanov}}{{Komissarov} \&
  {Gubanov}}{1994}]{KomissarovGubanov94}
{Komissarov} S.~S.,  {Gubanov} A.~G., 1994, \aap, 285, 27

\bibitem[\protect\citeauthoryear{{Mazzotta} et~al.}{{Mazzotta}
  et~al.}{2002}]{Mazzotta}
{Mazzotta} P., {Kaastra} J.~S., {Paerels} F.~B., {Ferrigno} C., {Colafrancesco}
  S., {Mewe} R.,  {Forman} W.~R., 2002, \apjl, 567, L37

\bibitem[\protect\citeauthoryear{{McNamara}, {O'Connell}, \&
  {Bregman}}{{McNamara} et~al.}{1990}]{McNamara90}
{McNamara} B.~R., {O'Connell} R.~W.,  {Bregman} J.~N., 1990, \apj, 360, 20

\bibitem[\protect\citeauthoryear{{McNamara} et~al.}{{McNamara}
  et~al.}{2000}]{McNamaraHydra00}
{McNamara} B.~R. et~al., 2000, \apjl, 534, L135

\bibitem[\protect\citeauthoryear{{Owen}, {Eilek}, \& {Kassim}}{{Owen}
  et~al.}{2000}]{Owen00M87}
{Owen} F.~N., {Eilek} J.~A.,  {Kassim} N.~E., 2000, \apj, 543, 611

\bibitem[\protect\citeauthoryear{{Owen} \& {Ledlow}}{{Owen} \&
  {Ledlow}}{1997}]{Owen&Ledlow97}
{Owen} F.~N.,  {Ledlow} M.~J., 1997, \apjs, 108, 41

\bibitem[\protect\citeauthoryear{{Pedlar} et~al.}{{Pedlar}
  et~al.}{1990}]{Pedlar90}
{Pedlar} A., {Ghataure} H.~S., {Davies} R.~D., {Harrison} B.~A., {Perley} R.,
  {Crane} P.~C.,  {Unger} S.~W., 1990, \mnras, 246, 477

\bibitem[\protect\citeauthoryear{{Perley}, {Dreher}, \& {Cowan}}{{Perley}
  et~al.}{1984}]{PerleyCygnusA}
{Perley} R.~A., {Dreher} J.~W.,  {Cowan} J.~J., 1984, \apjl, 285, L35

\bibitem[\protect\citeauthoryear{{Reiprich} \& {B{\" o}hringer}}{{Reiprich} \&
  {B{\" o}hringer}}{2002}]{Reiprich}
{Reiprich} T.~H.,  {B{\" o}hringer} H., 2002, \apj, 567, 716

\bibitem[\protect\citeauthoryear{{Reynolds} et~al.}{{Reynolds}
  et~al.}{1996}]{ReynoldsACFjets}
{Reynolds} C.~S., {Fabian} A.~C., {Celotti} A.,  {Rees} M.~J., 1996, \mnras,
  283, 873

\bibitem[\protect\citeauthoryear{{Rizza} et~al.}{{Rizza}
  et~al.}{2000}]{Rizza99}
{Rizza} E., {Loken} C., {Bliton} M., {Roettiger} K., {Burns} J.~O.,  {Owen}
  F.~N., 2000, \aj, 119, 21

\bibitem[\protect\citeauthoryear{{Sanders} \& {Fabian}}{{Sanders} \&
  {Fabian}}{2002}]{SandersCent02}
{Sanders} J.~S.,  {Fabian} A.~C., 2002, \mnras, 331, 273

\bibitem[\protect\citeauthoryear{{Schmidt}, {Fabian}, \& {Sanders}}{{Schmidt}
  et~al.}{2002}]{Schmidt2002}
{Schmidt} R.~W., {Fabian} A.~C.,  {Sanders} J.~S., 2002, \mnras, 337, 71

\bibitem[\protect\citeauthoryear{{Slee} \& {Reynolds}}{{Slee} \&
  {Reynolds}}{1984}]{SleeReynolds84}
{Slee} O.~B.,  {Reynolds} J.~E., 1984, Proceedings of the Astronomical Society
  of Australia, 5, 516

\bibitem[\protect\citeauthoryear{{Slee} et~al.}{{Slee} et~al.}{2001}]{Slee01}
{Slee} O.~B., {Roy} A.~L., {Murgia} M., {Andernach} H.,  {Ehle} M., 2001, \aj,
  122, 1172

\bibitem[\protect\citeauthoryear{{Smith} et~al.}{{Smith}
  et~al.}{2002}]{SmithCygnusA2002}
{Smith} D.~A., {Wilson} A.~S., {Arnaud} K.~A., {Terashima} Y.,  {Young} A.~J.,
  2002, \apj, 565, 195

\bibitem[\protect\citeauthoryear{{Stefanachi}, {Venturi}, \&
  {Dallacasa}}{{Stefanachi} et~al.}{2002}]{Stefanachi02}
{Stefanachi} F., {Venturi} T.,  {Dallacasa} D., 2002, in Proceedings of the 6th
  EVN Symposium, p. 147

\bibitem[\protect\citeauthoryear{{Taylor}, {Barton}, \& {Ge}}{{Taylor}
  et~al.}{1994}]{TaylorA4059}
{Taylor} G.~B., {Barton} E.~J.,  {Ge} J., 1994, \aj, 107, 1942

\bibitem[\protect\citeauthoryear{{Taylor}, {Fabian}, \& {Allen}}{{Taylor}
  et~al.}{2002}]{TaylorCent02}
{Taylor} G.~B., {Fabian} A.~C.,  {Allen} S.~W., 2002, \mnras, 334, 769

\bibitem[\protect\citeauthoryear{{Taylor} et~al.}{{Taylor}
  et~al.}{1990}]{TaylorHydra90}
{Taylor} G.~B., {Perley} R.~A., {Inoue} M., {Kato} T., {Tabara} H.,  {Aizu} K.,
  1990, \apj, 360, 41

\bibitem[\protect\citeauthoryear{{Ter{\" a}sranta} et~al.}{{Ter{\" a}sranta}
  et~al.}{2001}]{Terasranta01}
{Ter{\" a}sranta} H., {Urpo} S., {Wiren} S.,  {Valtonen} M., 2001, \aap, 368,
  431

\bibitem[\protect\citeauthoryear{{Voigt} \& {Fabian}}{{Voigt} \&
  {Fabian}}{2004}]{Voigt04}
{Voigt} L.~M.,  {Fabian} A.~C., 2004, in prep

\bibitem[\protect\citeauthoryear{{Zhao} et~al.}{{Zhao} et~al.}{1993}]{Zhao93}
{Zhao} J., {Sumi} D.~M., {Burns} J.~O.,  {Duric} N., 1993, \apj, 416, 51

\end{thebibliography}

\section*{Appendix} \label{appendix}\nonumber

The bubble timescales and derived powers are
presented in Table \ref{lum_t_table}.

\begin{table*} \centering \caption{\scshape Timescales and Powers}

\begin{tabular}{l l l l l l l l l l l l l}
\hline
\hline
Cluster& Lobe$^{(1)}$&$t_{\rm CMB}^{(2)} $& $t_{\rm sound}$&$t_{\rm shock}$ & $t_{\rm buoy}$
&$t_{\rm refill}$&$t_{\rm min}$ & $t_{\rm max}$ & $\mathcal{P}_{\rm sound}^{(3)}$ & $\mathcal{P}_{\rm shock}$ &
$\mathcal{P}_{\rm buoy}$ & $\mathcal{P}_{\rm refill}$\\ 
& &$10^7 \rm{yr}$&$10^7 \rm{yr}$ &$10^7 {\rm yr}$ &$10^7 {\rm yr}$ &$10^7
{\rm yr}$&$10^7 {\rm yr}$&$10^7 {\rm yr}$&$10^{43}{\rm ergs}^{-1}$
&$10^{43}{\rm ergs}^{-1}$ &$10^{43}{\rm ergs}^{-1}$ &$10^{43}{\rm ergs}^{-1}$\\ 

\hline
A133    & Rel, R&  12.38 &  13.9  &    -  &  1.99    &   6.93 & 1.26  & 1.22  &   4.74  &   -   & 33.1  & 9.52\\   
\\
A2052   & N, X  &  13.21 &   4.11 &    -  &  0.39    &   1.49 & 0.68  & 0.26  &   1.80  &   -   & 18.8  & 4.98\\   
        & S, X  &  13.21 &   6.16 &    -  &  0.65    &   2.44 & 1.03  & 0.43  &   4.06  &   -   & 38.6  & 10.2\\   
\\
A2199   & E, R  &  13.36 &   4.14 &    -  &  1.98    &   5.30 & 0.50  & 0.93  &   3.20  &   -   & 6.70  & 2.50\\   
        & W, R  &  13.36 &   5.04 &    -  &  1.91    &   5.82 & 0.58  & 1.03  &   4.47  &   -   & 11.8  & 3.86\\   
        & I-E, R&  13.36 &   0.30 &    -  &  1.54    &   1.37 & 0.05  & 0.24  &   0.05  &   -   & 0.01  & 0.01\\   
        & I-W, R&  13.36 &   0.45 &    -  &  1.26    &   1.68 & 0.075 & 0.30  &   0.12  &   -   & 0.04  & 0.03\\
\\
A4059   & N, R  &  12.65 &   1.66 &    -  &  0.25    &   0.71 & 0.15  & 0.13  &   0.12  &   -   & 0.80  & 0.29\\   
        &S, R   &  12.65 &   1.19 &    -  &  0.18    &   0.74 & 0.23  & 0.13  &   0.58  &   -   & 3.77  & 0.93\\   
\\
Centaurus& E, R &  14.19 &   0.70 &    -  &  0.31    &   1.02 & 0.091 & 0.18  &   0.02  &   -   & 0.05  & 0.01\\
        &E, R   &  14.19 &   0.81 &    -  &  0.46    &   1.20 & 0.073 & 0.21  &   0.01  &   -   & 0.03  & 0.01\\
\\
Cygnus A& E, R  &  12.43 &   6.53 &    -  &  5.09    &  11.7  & 0.85  & 2.06  &  56.3   &   -   & 72.2  & 31.5\\   
        & W, R  &  12.43 &   6.23 &    -  &  4.41    &  10.1  & 0.77  & 1.79  &  77.5   &   -   & 109   & 47.7\\   
\\
Hydra A & N, R  &  12.57 &   5.04 &    -  &  1.25    &   3.44 & 0.35  & 0.61  &   1.98  &   -   & 8.00  & 2.90\\   
        & S, R  &  12.57 &   4.16 &    -  &  2.13    &   3.87 & 0.28  & 0.68  &   1.16  &   -   & 2.26  & 1.24\\   
\\
M84     &N, R   &  14.48 &   1.12 &    -  &  0.40    &   0.74 & 0.23  & 0.13  &   0.05  &   -   & 0.21  & 0.08\\   
        &S, R   &  14.48 &   2.14 &    -  &  0.56    &   1.05 & 0.23  & 0.19  &   0.03  &   -   & 0.16  & 0.06\\   
\\
M87     & Bud, R&  14.45 &   0.46 &    -  &  0.55    &   0.89 & 0.077 & 0.16  &   0.20  &   -   & 0.16  & 0.10\\
\\
MKW3s   & S, R  &  12.83 &   3.11 &    -  &  5.52    &   5.60 & 0.72  & 0.99  &   3.22  &   -   & 1.81  & 1.79\\   
\\
Perseus & N, X  &  13.87 &   2.18 &  3.20 &  1.19    &   4.50 & 0.36  & 0.80  &   5.70  & 3.87  & 10.4  & 2.75\\   
        & S, X  &  13.87 &   2.15 &  2.90 &  1.30    &   4.90 & 0.36  & 0.87  &   7.84  & 5.81  & 13.0  & 3.44\\   
\\
PKS1404 & E, X  &  13.72 &   1.02 &    -  &  0.40    &   0.74 & 0.12  & 0.13  &   0.03  &   -   & 0.07  & 0.04\\   
        & W, X  &  13.72 &   0.63 &    -  &  0.26    &   0.53 & 0.14  & 0.094 &   0.06  &   -   & 0.14  & 0.07\\ 

\hline
\end{tabular} \label{lum_t_table}
\begin{quote}
{\scshape Notes:}

(1) The codes for the Lobes are N---Northern, S---Southern, E---Eastern,
    W---Western, X---sizes from X-ray image, R---sizes from Radio image,
    I-E/W---Inner lobes in A2199, Bud---as described in
    \citet{FormanM87}, Rel---Relic source as described in
    \citet{FujitaSarazin02}.

(2) The timescale for the bubble calculated from the magnetic field
    which produces the same energy density as that of the CMB at the
    redshift of the cluster for electrons radiating at $1\ghz$.

(3) The power is the PV/t work only, with $\frac{\gamma}{\gamma-1}$ not
    accounted for.  Therefore, for a fully relativistic plasma the
    values for the powers need to be multiplied by four, and for a
    non-relativistic plasma, by $5/2$.

\end{quote}
\end{table*}

\end{document}